\documentclass[fleqn,usenatbib]{mnras}
\usepackage{newtxtext,newtxmath}
\usepackage{soul}
\usepackage{float}
\usepackage[T1]{fontenc}
\usepackage{gensymb}
\usepackage{afterpage}
\DeclareRobustCommand{\VAN}[3]{#2}
\let\VANthebibliography\thebibliography
\def\thebibliography{\DeclareRobustCommand{\VAN}[3]{##3}\VANthebibliography}
\usepackage{graphicx}	
\usepackage{amsmath}	
\usepackage{siunitx}
\usepackage{orcidlink}

\title[Angular clustering of MeerKLASS sources]{Tracing Large-scale Structure with the MeerKLASS On-the-Fly Survey: Angular Clustering of Radio Sources at 816 MHz}

\author[S. Paul et al.]{Sourabh Paul$^1$\thanks{sourabh.paul@gmail.com}\,\orcidlink{0000-0002-8671-2177},
Suman Chatterjee$^2$,
Keith Grainge$^1$,
Laura Wolz$^1$,
Aishrila Mazumder$^1$,
\newauthor
Steven Cunnington$^3$,
Sarvesh Mangla$^4$,
Joseph J. Mohr$^4$,
Mario G. Santos$^{2,5}$,
Oleg Smirnov$^{6,5,7}$,
\newauthor
Cyril Tasse$^{8,9}$
\\
\\
$^1$Jodrell Bank Centre for Astrophysics, Department of Physics \& Astronomy, The University of Manchester, Manchester M13 9PL, UK\\
$^2$Department of Physics and Astronomy, University of the Western Cape, Robert Sobukwe Road, Cape Town 7535, South Africa\\
$^3$Institute of Cosmology and Gravitation, University of Portsmouth, Portsmouth, PO1 2UP, United Kingdom\\
$^4$University Observatory, LMU Faculty of Physics, Scheinerstr. 1, 81679, Munich, Germany\\
$^5$South African Radio Astronomy Observatory (SARAO), Cape Town, 7925, South Africa\\
$^6$Centre for Radio Astronomy Techniques and Technologies (RATT), Department of Physics and Electronics, Rhodes University, Makhanda, 6140, South Africa\\
$^7$Institute for Radioastronomy, National Institute of Astrophysics (INAF IRA), Via Gobetti 101, 40129 Bologna, Italy\\
$^8$GEPI \& ORN, Observatoire de Paris, Université PSL, CNRS, 5 Place Jules Janssen, 92190 Meudon, France\\
$^9$Department of Physics \& Electronics, Rhodes University, PO Box 94, Grahamstown, 6140, South Africa\\
}
\date{Accepted XXX. Received YYY; in original form ZZZ}

\pubyear{\the\year{}}

\begin{document}
\label{firstpage}
\pagerange{\pageref{firstpage}--\pageref{lastpage}}
\maketitle

\begin{abstract}
We present the first measurement of the angular two-point correlation function \(w(\theta)\) of radio sources from the MeerKAT Large Area Synoptic Survey (MeerKLASS) UHF on-the-fly (OTF) continuum Data Release~1. DR1 provides interferometric Stokes-\(I\) imaging at a reference frequency of 816\,MHz over \(\sim 800~\mathrm{deg}^2\) within the DESI footprint. We detect a positive clustering signal over \(0.02^\circ \lesssim \theta \lesssim 10^\circ\). The measurement is stable to reasonable variations of the depth mask and flux threshold on intermediate and large scales. Modelling the intermediate-scale signal (\(0.112^\circ\le\theta\le 1.36^\circ\)) with a fixed-slope power law (\(\gamma=1.8\)) yields \(A(1^\circ)=(1.434\pm0.475)\times10^{-3}\), corresponding to \(\log_{10}A=-2.843^{+0.124}_{-0.175}\). We infer an effective large-scale bias by fitting \(\Lambda\)CDM projected-matter templates \(w_{\rm DM}(\theta)\) computed with \textsc{CAMB} and Limber projection, including an integral-constraint correction evaluated from random--random weights. Using two bracketing T-RECS redshift-distribution priors, we obtain \(b_{\rm eff}=1.998\pm0.350\) (AGN prior) and \(b_{\rm eff}=1.530\pm0.265\) (TOTAL prior), demonstrating that the dominant modelling uncertainty arises from \(N(z)\). As a derived summary we Limber-invert the power-law amplitude to obtain \(r_0=6.18\pm1.13\) and \(5.59\pm1.02~h^{-1}\mathrm{Mpc}\) for the AGN and TOTAL priors, respectively. These results establish MeerKLASS UHF DR1 as a new wide-area, intermediate-frequency dataset for radio-continuum clustering. As MeerKLASS expands and overlapping optical/IR spectroscopy provides improved redshift calibration, future releases will enable population-split clustering and bias evolution measurements with substantially reduced modelling uncertainty.
\end{abstract}


\section{Introduction}
\label{sec:intro}

The spatial distribution of radio sources on the sky encodes key information about the growth of structure and the connection between galaxies and their underlying dark-matter haloes \citep{Kaiser_1984, Mo_White_1996, Cooray_Seth_2002, Camera_2012, Ferramacho_2014, jarvis_2015_tvdtx-ah378}. In the standard picture, radio-loud active galactic nuclei (AGN) and star-forming galaxies (SFGs) trace the matter density field in a biased manner, with the effective bias depending on luminosity, redshift, and the evolving mix of source populations \citep{Hale_2018,Wilman_2008,Magliocchetti_2017,Bonaldi_2019}. Measurements of two-point clustering therefore provide a route to constrain the typical halo environments of radio emitters, test phenomenological and semi-empirical models of the radio population, and deliver large-scale-structure constraints that are complementary to optical surveys \citep{Blake_Wall_2002, Overzier_2003, Raccanelli_2015}.

Wide-area radio continuum surveys are particularly powerful for clustering studies because they provide very large, relatively uniform source samples over thousands of square degrees, while being largely insensitive to dust extinction that can bias or obscure high-redshift populations in the optical \citep{Hao_2011,Cucciati_2012,Singh_2014,Saxena_2017}. Although optical surveys can provide direct redshift information, obtaining spectroscopic redshifts over comparably large sky areas and to sufficient depth is observationally expensive, while photometric redshifts introduce their own uncertainties and selection effects. Radio surveys therefore offer an efficient and complementary route to large-scale-structure measurements, especially for bright samples dominated by radio-loud AGN that extend to high redshift. Classic measurements using surveys such as FIRST \citep{becker1994} and NVSS \citep{condon1998} established the large-scale clustering of bright, AGN-dominated radio populations. Today, new-generation interferometers such as LOFAR \citep{vanHaarlem2013}, ASKAP \citep{DeBoer2009}, and MeerKAT \citep{jonas2009} have revolutionised the field by enabling higher-resolution analyses, pushing to significantly deeper flux limits, and improving the control of systematics over vast footprints. Beyond autocorrelation measurements, wide-area radio samples also underpin a growing set of cosmological applications through cross-correlations with other large-scale-structure tracers and the Cosmic Microwave Background (CMB), providing complementary constraints on structure growth on the largest scales. These include cosmic radio dipole, integrated Sachs--Wolfe measurements and CMB-lensing cross-correlations \citep{BlakeWall2002,RubartSchwarz2013,Bohme2025,Giannantonio2008,Planck_lensing_2014}.

Robust clustering measurements from radio-continuum catalogues are, however, non-trivial in practice. Unlike idealised galaxy samples, radio catalogues can exhibit (i) spatially varying completeness driven by position-dependent rms noise and deconvolution artefacts, (ii) complex source morphology in which a single physical radio galaxy may be resolved into multiple catalogue components \citep{Blake_Wall_2002,Cress_1996}, and (iii) the absence of direct redshift information for most sources, which complicates physical interpretation. Each of these effects can imprint artificial large-scale gradients or small-scale excess power if not explicitly modelled. A key requirement is therefore to construct a survey mask and random catalogues that accurately encode the footprint and depth variations \citep{Chakraborty_2020,Mazumder_2022,Hale_2024}. In wide-area analyses, it is also essential to propagate correlated uncertainties using an empirical covariance \citep{Norberg_2009}. 

Recent progress has been driven by surveys that either cover extremely large areas at low radio frequencies \citep{shimwell2022,Hale_2024}, or provide classic all-sky benchmarks at \(\sim 1.4\)~GHz \citep{condon1998,becker1994}, while newer wide-area surveys at a few GHz extend this landscape with higher angular resolution and multi-epoch coverage \citep{lacy2020}. Nevertheless, there remains strong value in establishing clustering measurements at intermediate \(\sim\)GHz frequencies with modern imaging fidelity, well-characterised depth variations, and overlap with contemporary optical spectroscopic programmes \citep{mcconnell2020,Hopkins2025}. This regime provides an important cross-check of frequency-dependent selection effects and enables cross-correlations with external large-scale-structure tracers, particularly in the southern hemisphere.

The MeerKAT radio telescope, a highly sensitive Square Kilometre Array (SKA) precursor in South Africa, provides an excellent platform for such wide-area continuum and large-scale-structure studies at \(\sim\)GHz frequencies \citep{jonas2009}. Its sensitivity and imaging fidelity make it well suited to producing high-quality wide-area continuum mosaics, particularly in the southern hemisphere where comparable depth--area combinations remain relatively scarce. The MeerKAT Large Area Synoptic Survey (MeerKLASS; \citealt{meerklass}) is a flagship commensal programme that exploits these capabilities by combining \ion{H}{i} intensity mapping \citep{Chang_2010,Masui_2013,Anderson_2018,Wolz_2022,Cunnington_2023,Paul_2023,CHIME_2025} with interferometric continuum imaging in the MeerKAT L and UHF bands \citep{Chatterjee_2025,Mangla_2025,Paul_2025}. The full MeerKLASS programme targets wide, contiguous coverage over a large fraction of the southern sky (outside the Galactic plane), with an overall goal of surveying \(\sim10{,}000~\mathrm{deg}^2\) in \(\sim2500\) hours of night-time observing. This is enabled by a constant-elevation \emph{on-the-fly} strategy in which MeerKAT scans continuously in azimuth at a fixed rate, providing efficient mapping, a stable point-spread function, and well-behaved mosaicking properties. The cross-linked rising and setting scans improve sky coverage uniformity and \(uv\)-coverage, helping to suppress imaging artefacts. 

The first public data release of the MeerKLASS UHF OTF continuum survey (\citealt{Paul_2025}, hereafter DR1) demonstrates the viability of this approach and delivers deep, high-fidelity continuum imaging across an \(\sim 800~\mathrm{deg}^2\) region within the Dark Energy Spectroscopic Instrument (DESI; \citealt{DESI}) footprint, together with a validated source catalogue. DR1 reaches an rms of \(\sim 35~\mu\mathrm{Jy~beam^{-1}}\) in the deepest overlap regions at a central frequency of 816\,MHz. These characteristics make MeerKLASS an attractive new dataset for clustering analyses at \(\sim\)GHz frequencies. As MeerKLASS expands in area and depth in future releases, the combination of wide-area continuum imaging and overlapping optical spectroscopy will enable population-split clustering, redshift-calibrated bias evolution, and precision cross-correlations with external tracers.

In this paper we present the first measurement of the angular clustering of DR1 radio sources and infer the effective large-scale bias of a conservative, well-controlled sample. Our analysis proceeds in two stages. First, we measure the angular two-point correlation function $w(\theta)$ using a mask- and depth-controlled selection and a standard Landy--Szalay estimator, with uncertainties estimated from jackknife resampling. Second, we interpret the measured clustering amplitude by comparing to a $\Lambda$CDM template prediction for the projected matter clustering, computed for a set of redshift-distribution priors $N(z)$ derived from semi-empirical radio simulations. The outcome is a constraint on the effective bias of the sample together with a quantitative assessment of the dominant modelling uncertainties. This paper is organised as follows. In \autoref{sec:data} we summarise the MeerKLASS DR1 dataset, define the clustering sample, and describe the survey mask and random catalogues. In \autoref{sec:wtheta} we outline the estimator, binning choices, covariance estimation, and integral-constraint treatment used to obtain $w(\theta)$. In \autoref{sec:results}, we present the measured angular correlation function. The power-law characterisation of $w(\theta)$ is discussed in \autoref{sec:powerlaw}. In \autoref{sec:bias}, we interpret the measurement in terms of large-scale bias using projected matter templates and redshift-distribution priors, and in \autoref{sec:comparison}, we compare our results with previous radio clustering studies. Finally, in \autoref{sec:conclusions} we summarise our main conclusions. Throughout this work, we assume a spatially flat $\Lambda$CDM cosmology with parameters consistent with the \textit{Planck} 2018 results \citep{Planck2018}.

\section{Data and sample definition}
\label{sec:data}

\subsection{MeerKLASS UHF DR1 overview}
\label{subsec:dr1_overview}

This work uses the first public interferometric data release from the MeerKLASS UHF OTF continuum survey \citep{Paul_2025}. DR1 is based on eight OTF interferometric scan blocks (four rising and four setting) totalling $\simeq 12$ hours of usable observing time and covering $\sim 800~\mathrm{deg}^2$ within the DESI footprint (\autoref{fig:sky_footprint} top panel). In OTF mode, MeerKAT scans continuously in azimuth at constant elevation at a scan speed of $7$ arcmin~s$^{-1}$, with visibilities recorded at 2\,s cadence. The rising/setting strategy produces cross-linked scan tracks in equatorial coordinates, yielding enhanced depth where multiple blocks overlap and shallower coverage towards edges covered by only one (or a few) passes. As a result, the final DR1 footprint exhibits an intrinsically non-uniform depth pattern that must be explicitly encoded in any clustering analysis to avoid artificial large-scale gradients in the source density field.

For the interferometric continuum products, DR1 provides Stokes-$I$ images referenced to a central frequency of 816\,MHz. Imaging is performed using a visibility-domain mosaicking approach with a customised \textsc{DDFacet} workflow designed for fast OTF data \citep{tasse2018}. The survey region is partitioned into 89 overlapping tiles of size $3.2^\circ \times 3.2^\circ$ with $\simeq 0.1^\circ$ overlap between adjacent tiles. For each tile, measurement sets within a larger $4^\circ \times 4^\circ$ selection region are imaged jointly, producing a wide ($\sim 7^\circ$) deconvolution field from which the central $3.2^\circ \times 3.2^\circ$ science cutout is retained. All tiles share a common pixel scale of 3\,arcsec, and the delivered restoring beam is typically $\sim 32\arcsec \times 17\arcsec$ at 816\,MHz. Although the detailed $uv$ coverage (and hence the PSF) varies across the footprint because different sky patches receive different numbers of rising/setting passes, the DR1 processing delivers a stable, survey-wide resolution that is well-characterised on a per-tile basis.

A key consideration for clustering estimation is the spatially varying survey depth. DR1 therefore provides a survey-scale rms noise map constructed from the residual mosaics. The noise is estimated using sigma-clipped statistics in a sliding window of $100\times 100$ pixels ($5^\prime \times 5^\prime$ at 3\,arcsec resolution), stepped by 50 pixels across the full residual mosaic. This map traces the expected depth enhancement in overlap regions and highlights localised ``high-rms islands'' associated with dynamic-range limitations around very bright or extended sources. In our clustering analysis, this rms map is used to define conservative depth cuts and to construct random catalogues that follow the same angular selection function as the data (\autoref{fig:sky_footprint} bottom panel).

Source detection in DR1 is performed tile-by-tile with \textsc{PyBDSF} \citep{mohan2015pybdsf}, and the resulting catalogues are merged using an explicit cross-tile de-duplication scheme. The primary DR1 source catalogue (SRL) contains 95{,}483 unique sources, and the associated Gaussian-component catalogue (GAUL) contains 115{,}328 fitted components. For clustering, the SRL catalogue provides (i) integrated flux densities (in Jy), (ii) peak fluxes (in Jy~beam$^{-1}$), (iii) per-source noise estimates (e.g. \texttt{Isl\_rms}), and (iv) morphology/fit metadata (e.g. \texttt{S\_Code}, \texttt{N\_Gaus}) that enable conservative ``compact-only'' selections to mitigate multi-component blending on small angular scales. The combination of a validated source catalogue, per-tile beam metadata, and a survey-scale rms map makes DR1 a self-contained dataset for wide-area angular clustering measurements, provided the depth variations are explicitly modelled through the mask and randoms (as described below).

\subsection{Clustering sample}
\label{subsec:sample}

From the DR1 SRL catalogue we construct a conservative clustering sample designed to minimise spatial fluctuations in the observed source density driven by depth variations, reduce contamination from multi-component radio galaxies on small angular scales, and ensure that the adopted flux-density threshold lies safely above the effective detection limit across the analysed footprint. Throughout, flux densities are taken from the SRL integrated (total) flux column (in Jy) and are quoted at the DR1 reference frequency of 816\,MHz.

\begin{figure*}
    \centering
    \includegraphics[width=\textwidth, trim=0 30 0 30, clip]{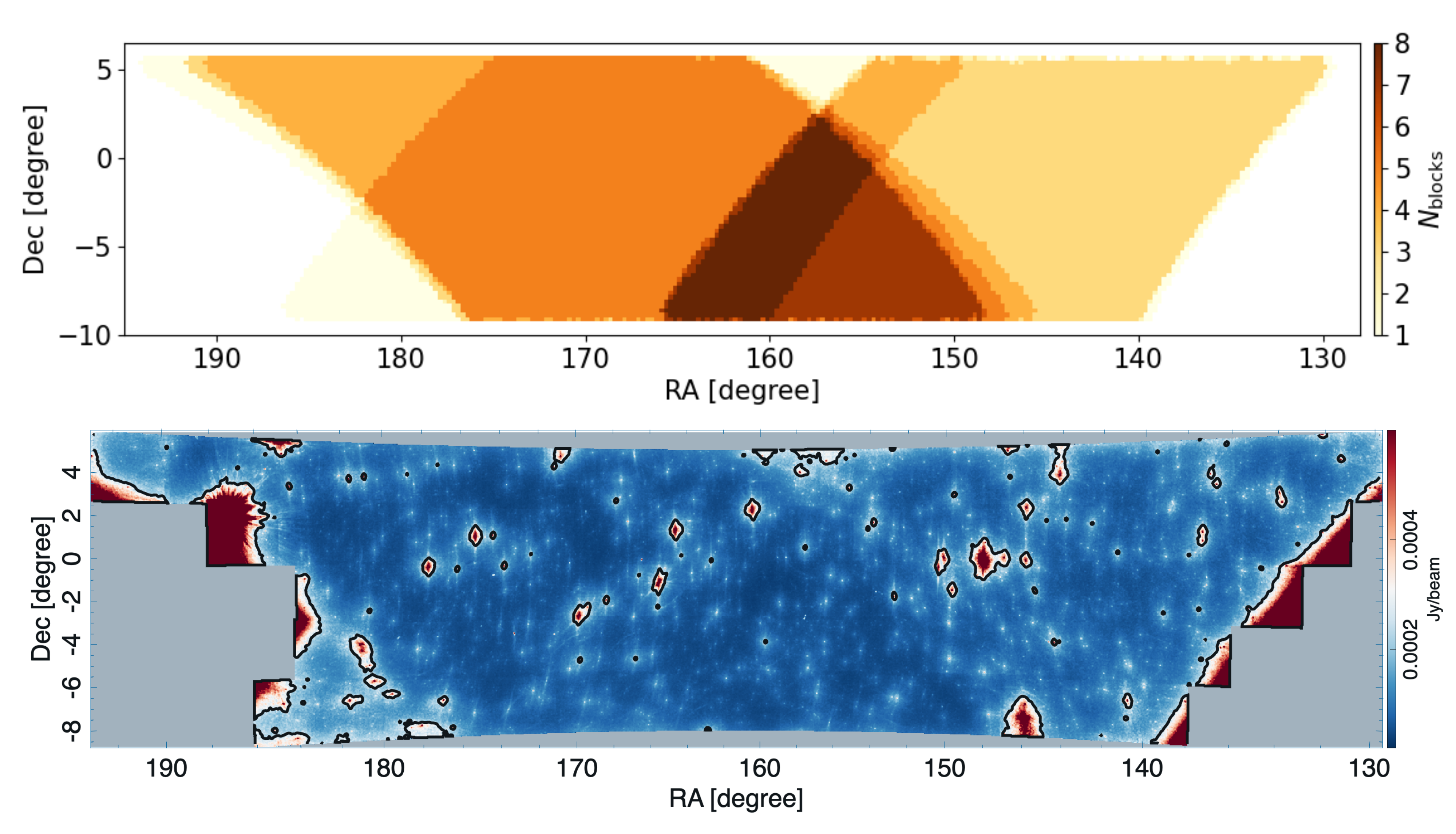}
    \caption{Survey footprint, depth variations, and clustering mask for MeerKLASS UHF DR1. \textbf{Top:} pixelised coverage map showing the number of OTF interferometric scan blocks contributing to each sky position (\(N_{\rm blocks}=1\)–8). The deepest region occurs where rising and setting passes overlap, while coverage (and hence sensitivity) decreases towards the survey edges that are sampled by fewer blocks. \textbf{Bottom:} local rms noise map at 816\,MHz, \(\sigma_{\rm local}\) (Jy\,beam\(^{-1}\)), derived from the DR1 residual mosaic. The large-scale depth pattern broadly follows the expected \(\sigma_{\rm local}\propto N_{\rm blocks}^{-1/2}\) behaviour, with localised high-rms ``islands'' around very bright and/or extended sources where dynamic-range limitations inflate the residual-based rms estimate. Black contour marks the adopted depth boundary \(\sigma_{\rm local}=250~\mu\mathrm{Jy\,beam^{-1}}\). This binary mask defines the angular selection function used to generate random catalogues and to measure \(w(\theta)\).}
    \label{fig:sky_footprint}
\end{figure*}

Our fiducial selection is defined by the following cuts:
\begin{itemize}
    \item \textbf{Flux-density threshold.} We require \(S_{816} > S_{\rm lim}\) with \(S_{\rm lim}=2~\mathrm{mJy}\). This choice is conservative with respect to the DR1 injection--recovery completeness analysis (Fig.~14 of \citealt{Paul_2025}), in which the survey-wide median completeness reaches \(\sim 50\%\) at \(\simeq 0.6\)~mJy and approaches \(\sim 90\%\) by \(\simeq 1.5\)~mJy, with central overlap tiles being systematically deeper and a tail of poorer-performing tiles at faint fluxes. In the clustering analysis we further suppress any residual completeness-driven surface-density fluctuations by applying an explicit depth-controlled mask (discussed below), which removes the noisiest regions. These excluded areas occur predominantly near the survey edges (lower sky redundancy) and in localised high-rms ``islands'' around very bright/extended sources, and therefore exhibit degraded detection performance (see the bottom panel of \autoref{fig:sky_footprint} and the rms-map discussion in \citealt{Paul_2025}).

    \item \textbf{Depth-controlled footprint.} We restrict the analysis to positions where the local rms noise from the DR1 survey-scale rms map satisfies \(\sigma_{\rm local} \le \sigma_{\rm lim}\) with \(\sigma_{\rm lim}=250~\mu\mathrm{Jy\,beam^{-1}}\). This depth cut trims the high-noise tail (typically occurring near footprint edges and in localised high-rms regions around very bright/extended sources) and provides a well-defined angular selection function for constructing random catalogues.

    \item \textbf{Compact-only (single-component) sample.} To mitigate the well-known small-scale excess produced when a single physical radio galaxy is split into multiple catalogue components, we adopt a compact, single-component selection using SRL metadata. Unless stated otherwise, our fiducial results use sources satisfying \texttt{S\_Code = 'S'} and \texttt{N\_Gaus = 1}. This choice yields a cleaner two-halo regime for power-law fitting and for the template-based bias inference presented later. For validation, we also consider an ``all-sources'' sample (i.e. without the compactness restriction) as a dedicated stress test of morphology-driven small-scale systematics.
\end{itemize}

\begin{figure*}
    \centering
    \includegraphics[width=\textwidth, trim=0 0 0 0, clip]{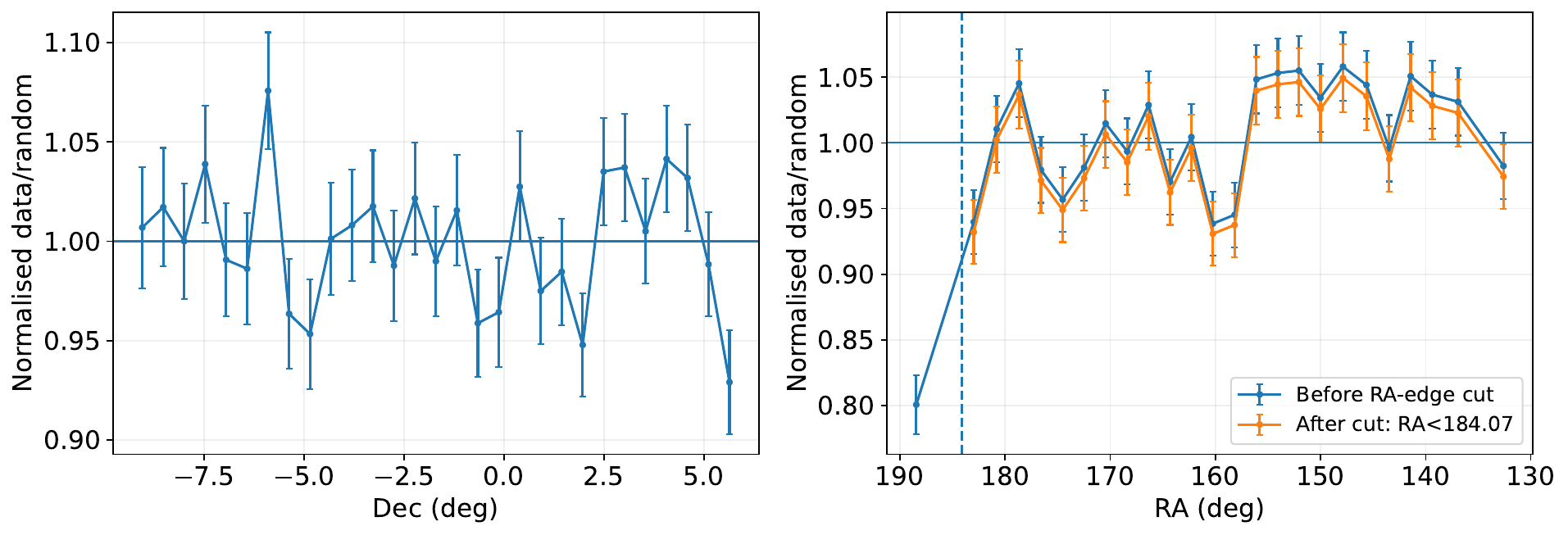}
    \caption{Uniformity diagnostics for the fiducial clustering selection. \textbf{Left:} ratio of the normalised data and random surface densities, \(R \equiv n_{\rm data}/n_{\rm rand}\), in declination using 30 equal-width bins (normalised such that the mean ratio is unity). \textbf{Right:} the same ratio in right ascension using 30 equal-area bins, where the bin boundaries are defined by RA-quantiles of the random catalogue so that each bin contains (approximately) the same expected number of random points under the survey mask. Blue points show the RA profile before applying the RA-edge cut; orange points show the profile after imposing \(\mathrm{RA}\le 184.07^\circ\) (vertical dashed line). Error bars denote approximate \(1\sigma\) Poisson uncertainties on the ratio (propagated from the binned data and random counts). The deficit in the final RA bin motivates the conservative RA-edge cut, after which the RA profile is consistent with unity within uncertainties.}
    \label{fig:uniformity_diagnostics}
\end{figure*}

\subsection{Survey mask and random catalogues}
\label{subsec:mask_random}

A robust angular clustering measurement requires an explicit description of the \emph{angular selection function}: the probability that a source meeting the flux and morphology cuts would be included at a given sky position. For DR1, the dominant selection effects are geometric (the finite mosaic footprint) and depth-related (spatially varying rms noise driven by the OTF overlap pattern and localised high-rms regions). We therefore construct a \emph{binary} clustering mask using the DR1 survey-scale rms map (\autoref{subsec:dr1_overview}), such that only sky positions within the valid mosaic footprint and satisfying the adopted depth criterion are included in the analysis.

Specifically, the fiducial mask is defined by the intersection of:

\begin{itemize}
    \item \textbf{Footprint geometry:} the union of all valid (finite) pixels in the DR1 UHF mosaic, excluding regions outside the imaged area.
    \item \textbf{Depth (rms) mask:} a threshold on the local rms noise, retaining only pixels with \(\sigma_{\rm local} \le \sigma_{\rm lim}\) (with \(\sigma_{\rm lim}=250~\mu\mathrm{Jy\,beam^{-1}}\) for the fiducial analysis).
\end{itemize}

This binary depth mask is intentionally conservative: it removes the high-noise tail while providing a simple selection function that can be reproduced exactly when generating random catalogues.

We generate an initial random catalogue that realises the angular selection defined by the geometric footprint and the rms depth mask. Candidate sky positions are drawn uniformly in right ascension \(\alpha\) and uniformly in \(\sin\delta\) (to ensure uniform sampling in solid angle) within a bounding box enclosing the survey region. Each candidate point is then filtered through the mask by evaluating \(\sigma_{\rm local}(\hat{\mathbf{n}})\) at its sky position using the rms-map WCS; to avoid discretisation artefacts from nearest-pixel assignment we sample the rms field using bilinear interpolation over the four nearest pixels. A point is accepted if it (i) falls within the valid mosaic footprint and (ii) satisfies \(\sigma_{\rm local} \le \sigma_{\rm lim}\). The resulting accepted random catalogue is therefore uniform \emph{within} the binary depth mask and, by construction, shares the same baseline angular selection as the data.

To validate the angular selection function encoded by the mask and random catalogues, we compare the spatial distribution of the clustering sample to that of the randoms. \autoref{fig:uniformity_diagnostics} shows the ratio
\(R \equiv n_{\rm data}/n_{\rm rand}\), where \(n_{\rm data}\) and \(n_{\rm rand}\) are counts in each bin normalised by the total counts in the respective catalogues, so that \(R=1\) is expected for a uniform selection. We compute \(R\) in 30 equal-width declination bins (left panel) and in 30 equal-area right-ascension bins (right panel). The RA bin boundaries are defined using quantiles of the random catalogue, ensuring comparable expected counts per bin under the survey mask. We find that the declination profile is consistent with unity within uncertainties, while the most extreme RA bin at the survey edge exhibits a significant deficit relative to the rest of the footprint. We therefore adopt a conservative RA-edge cut, \(\mathrm{RA} \le 184.07^\circ\), which removes this final RA bin and yields a substantially flatter \(n_{\rm data}/n_{\rm rand}\) profile across RA. This cut is applied consistently to both the data and random catalogues and is included as part of the fiducial clustering mask. In the right panel, blue points show the RA profile before applying the RA-edge cut, while orange points show the profile after imposing the cut. Error bars are estimated from Poisson counting statistics for the (unnormalised) bin counts, propagated to the ratio.

After applying the fiducial flux, depth, compactness, and RA-edge selections, the final clustering catalogue contains \(N_{\rm data}=39{,}483\) sources.

\section{Measuring the angular correlation function}
\label{sec:wtheta}

\subsection{Two-point correlation functions in real and angular space}
\label{subsec:two_point_review}

The clustering of discrete sources on the sky is commonly quantified using two-point correlation functions, which measure the excess probability (relative to a random distribution) of finding pairs separated by a given distance. In three-dimensional comoving space, the two-point correlation function \(\xi(r)\) is defined such that the joint probability of finding objects in two volume elements \(dV_1\) and \(dV_2\) separated by \(r\) is \citep{Peebles1980}
\begin{equation}
dP = \bar{n}^2\,\left[1+\xi(r)\right]\,dV_1\,dV_2,
\end{equation}
where \(\bar{n}\) is the mean number density. For purely angular catalogues (as in radio-continuum surveys lacking individual redshifts), the analogous statistic is the angular correlation function \(w(\theta)\), defined by \citep{Peebles1980}
\begin{equation}
dP = \bar{n}_\Omega^2\,\left[1+w(\theta)\right]\,d\Omega_1\,d\Omega_2,
\label{eq:wtheta_def}
\end{equation}
where \(d\Omega_1\) and \(d\Omega_2\) are infinitesimal solid-angle elements on the sky around two directions separated by an angle \(\theta\), and \(\bar{n}_\Omega\) is the mean surface density of sources (number per unit solid angle). For an unclustered (Poisson) distribution on the sky, \(w(\theta)=0\) in expectation and any measured signal fluctuates around zero due to finite sampling and survey geometry \citep{Peebles1980}. In contrast, clustered galaxy populations typically exhibit \(w(\theta)>0\) over a broad range of angular scales. Empirically, \(w(\theta)\) is often well described by a power law on intermediate scales,
\begin{equation}
w(\theta) \approx A\left(\frac{\theta}{1^\circ}\right)^{1-\gamma}.
\label{w_theta_eq}
\end{equation}
Here, $A$ denotes the dimensionless amplitude of the clustering evaluated at $1^\circ$, and $\gamma$ is the power-law index which takes typical values of $\gamma \simeq 1.7$--$1.9$ in characteristic radio and optical samples. \citep{Groth1977,Peebles1980,Overzier_2003}. Deviations are expected at very small separations (e.g. multi-component sources and one-halo contributions) and on the largest scales where survey systematics and integral-constraint effects can become important \citep{Blake_Wall_2002,Overzier_2003,Hale_2024}.

The angular statistic \(w(\theta)\) is related to the underlying three-dimensional clustering through a projection along the line of sight, weighted by the redshift distribution \(N(z)\). Under the Limber approximation \citep{Limber1953}, \(w(\theta)\) can be expressed as a weighted integral over the matter power spectrum (or equivalently over \(\xi\)), making the interpretation of the measured amplitude sensitive to the assumed \(N(z)\). In this work we therefore first present a direct measurement of \(w(\theta)\), and subsequently interpret it via \(\Lambda\)CDM template modelling using plausible \(N(z)\) priors (\autoref{sec:bias}).

\subsection{Estimator, randoms, and binning}
\label{subsec:estimator}

We estimate the angular two-point correlation function \(w(\theta)\), which quantifies the excess probability (relative to a random distribution) of finding two sources separated by an angle \(\theta\) on the sky. For a discrete catalogue, \(w(\theta)\) is measured from pair counts using the minimum-variance Landy--Szalay estimator \citep{Landy_Szalay_1993},
\begin{equation}
w(\theta) \;=\; \frac{DD(\theta) - 2\,DR(\theta) + RR(\theta)}{RR(\theta)},
\label{eq:ls_estimator}
\end{equation}
where \(DD\), \(DR\), and \(RR\) are the \emph{normalised} data--data, data--random, and random--random pair counts in an angular bin around \(\theta\). The Landy--Szalay form suppresses boundary effects by explicitly accounting for the survey geometry through \(RR\) and \(DR\), and it reduces variance compared to simpler estimators by optimally combining the three pair counts.

A crucial input to \autoref{eq:ls_estimator} is a random catalogue that traces the same angular selection function as the data (\autoref{subsec:mask_random}). We generate random coordinates uniformly on the sphere by drawing \(\alpha\) uniformly and \(\sin\delta\) uniformly within a bounding box enclosing the survey region. Candidate positions are then filtered through the fiducial clustering mask described in \autoref{subsec:mask_random} (geometry + rms depth cut + RA-edge selection), yielding a random catalogue that is uniform \emph{within} the final mask and therefore shares the same angular selection function as the data. This random catalogue is used consistently in both the \(DR\) and \(RR\) pair counts in (\autoref{eq:ls_estimator}). Unless stated otherwise we use \(N_{\rm rand}=20\,N_{\rm data}\), which renders Poisson fluctuations in the \(DR\) and \(RR\) pair counts negligible compared to the data uncertainties. We further verify that our results are insensitive to the precise random density in appendix~\ref{app:random_convergence}. We note that some deep, small-area radio mosaics construct random catalogues via injection--recovery (injecting simulated sources into residual/noise maps and re-running the source finder) to capture subtle completeness variations near the detection threshold \citep{Chakraborty_2020,Mazumder_2022}. For our wide-area footprint and conservative flux cut, the RMS-based binary depth mask and the uniformity diagnostics provide a sufficiently accurate description of the angular selection for clustering.

Pair counts are computed with the \textsc{TreeCorr} package \citep{treecorr}, which efficiently evaluates separations on the sphere. We adopt \(N_{\rm bin}=20\) logarithmically spaced angular bins spanning \(0.02^\circ \le \theta \le 10^\circ\). This range is wide enough to probe the transition from sub-degree to multi-degree clustering while avoiding very small angles where multi-component sources can inflate the signal and very large angles where residual selection gradients can become more relevant. For each bin we quote the effective separation as the pair-weighted logarithmic mean,
\(\theta_{\rm eff}=\exp\langle \ln\theta\rangle\), as returned by \textsc{TreeCorr}.

\subsection{Covariance estimation}
\label{subsec:cov}

Measurements of \(w(\theta)\) in different angular bins are correlated because the same large-scale structures (and any residual survey systematics) contribute power across a range of separations. We therefore estimate the full covariance matrix using spatial jackknife resampling. Following standard practice, we partition the DR1 clustering footprint into \(N_{\rm JK}=30\) approximately equal-area, contiguous regions using \textsc{TreeCorr}'s internal patching scheme, and recompute \(w(\theta)\) \(N_{\rm JK}\) times, each time omitting one patch. The jackknife covariance is then
\begin{equation}
C_{ij} \;=\; \frac{N_{\rm JK}-1}{N_{\rm JK}}
\sum_{k=1}^{N_{\rm JK}}
\left[w_k(\theta_i)-\bar{w}(\theta_i)\right]
\left[w_k(\theta_j)-\bar{w}(\theta_j)\right],
\label{eq:jk_cov}
\end{equation}
where \(w_k\) denotes the correlation function measured with patch \(k\) removed and \(\bar{w}\) is the mean over all jackknife realisations. The corresponding \(1\sigma\) uncertainties are \(\sigma_i=\sqrt{C_{ii}}\), and we visualise bin-to-bin couplings using the correlation matrix \(\rho_{ij}=C_{ij}/\sqrt{C_{ii}C_{jj}}\).

When performing parameter inference using \(\chi^2\) fits that require the inverse covariance (precision matrix), the naive inverse of an estimated covariance can be biased when the number of resamplings is finite. We therefore apply the commonly used Hartlap scaling to the precision matrix \citep{Hartlap_2007, Alam_2017, Favole_2021},
\begin{equation}
\alpha_{\rm H} = \frac{N_{\rm JK}-N_{\rm bin}-2}{N_{\rm JK}-1},
\end{equation}
and use $C^{-1}_{\rm eff} = \alpha_{\rm H}\,C^{-1}$ for all parameter inference, where \(N_{\rm bin}\) is the number of angular bins included in the specific fit. We note that the Hartlap factor is derived under the assumption of independent covariance realisations; for jackknife tests it should be regarded as an approximate, conservative correction. We verify that our main conclusions are stable to reasonable variations of the jackknife partitioning (Appendix~\ref{app:jk_robustness}).

\subsection{Integral constraint}
\label{subsec:ic}

An angular correlation function estimated from a finite field is subject to the \emph{integral constraint} (IC): because the mean source density is inferred from the survey itself, the estimator effectively enforces that the correlation function averages to zero over the observed footprint. If the true correlation function is positive on large scales, this constraint suppresses the measured \(w(\theta)\) by an approximately constant offset, causing a small negative bias at all separations \citep{Peebles1980}.

For a given model prediction \(w_{\rm model}(\theta)\) and a specific survey mask, the IC can be written as the footprint-averaged value of the model \citep{Groth1977,Daddi_2001,Dalmasso_2024},
\begin{equation}
{\rm IC} \;=\; \frac{1}{\Omega^2}\int d\Omega_1 \int d\Omega_2\; w_{\rm model}(\theta_{12}),
\label{eq:ic_continuous}
\end{equation}
where \(\Omega\) is the solid angle of the survey and \(\theta_{12}\) is the angular separation between two directions within the mask. In practice, this double integral is most conveniently evaluated using the random catalogue that encodes the same angular selection function as the data. Approximating the integral by a discrete sum over random--random pairs gives
\begin{equation}
{\rm IC} \;=\; \frac{\sum_i RR(\theta_i)\,w_{\rm model}(\theta_i)}{\sum_i RR(\theta_i)},
\label{eq:ic_rr}
\end{equation}
where \(RR(\theta_i)\) are the (normalised) random--random pair counts in bin \(i\). This expression makes explicit that the IC depends on both the assumed model and the survey mask: deeper or more complex masks generally increase the weight of small separations and can change the magnitude of the correction.

When comparing a model to the measured correlation function we therefore subtract the IC from the theoretical prediction \citep{Peacock_1991},
\begin{equation}
w_{\rm pred}(\theta) \;=\; w_{\rm model}(\theta) - {\rm IC},
\label{eq:ic_apply}
\end{equation}
and fit \(w_{\rm pred}\) to the data using the full covariance. For the wide DR1 footprint and the conservative depth mask adopted here the IC is small, but we include it consistently in all \(\Lambda\)CDM template comparisons (\autoref{sec:bias}) and in any large-scale fits where it can otherwise bias the recovered amplitude.

\section{Results: Fiducial angular clustering}
\label{sec:results}

\subsection{Fiducial \(w(\theta)\) measurement}
\label{subsec:fiducial_wtheta}

\begin{figure*}
    \centering
    \includegraphics[width=\textwidth, trim=0 0 0 0, clip]{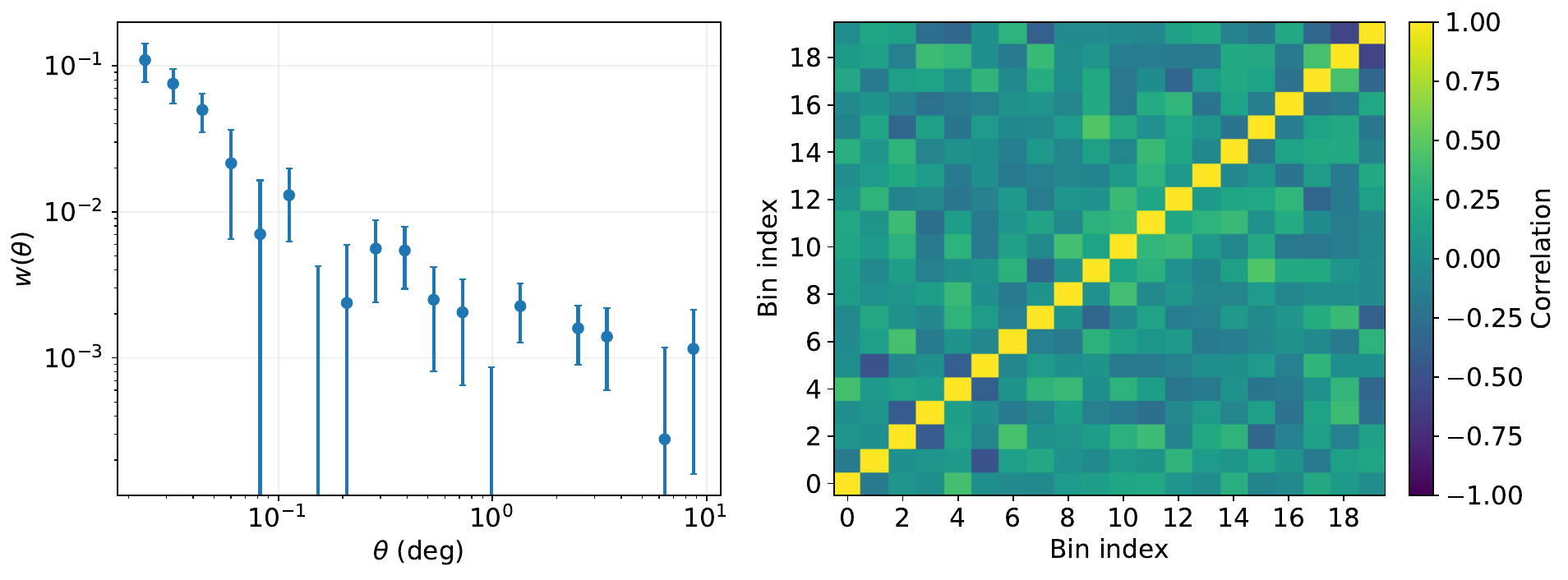}
    \caption{Fiducial angular clustering measurement and covariance. \textbf{Left:} measured angular correlation function \(w(\theta)\) in 20 logarithmically spaced bins spanning \(0.02^\circ\le\theta\le 10^\circ\); error bars show \(1\sigma\) uncertainties from the spatial jackknife covariance with \(N_{\rm JK}=30\) patches. \textbf{Right:} the corresponding jackknife \emph{correlation} matrix \(\rho_{ij}=C_{ij}/\sqrt{C_{ii}C_{jj}}\), illustrating the bin-to-bin correlations induced by large-scale structure and the finite survey footprint. This fiducial \(w(\theta)\) measurement forms the basis for the stability tests and the power-law and \(\Lambda\)CDM template fits presented in the following sections.}
    \label{fig:wtheta_cov}
\end{figure*}

\autoref{fig:wtheta_cov} (left) presents the fiducial measurement of the angular correlation function \(w(\theta)\) for the conservative, depth-controlled compact sample defined in \autoref{subsec:sample} and masked as described in \autoref{subsec:mask_random}. The measurement uses 20 logarithmically spaced bins spanning \(0.02^\circ\le\theta\le 10^\circ\), with uncertainties estimated from the \(N_{\rm JK}=30\) spatial jackknife covariance.

The measured correlation function is positive over a broad range of angular scales and declines smoothly with increasing separation, as expected for clustered radio sources projected over a wide redshift distribution. The largest amplitudes occur in the smallest-separation bins, although these scales are also the most susceptible to residual morphology-driven excess from fragmented multi-component sources. We therefore regard this measurement as the fiducial empirical \(w(\theta)\) estimate, while deferring a more detailed assessment of scale-dependent robustness and the choice of fitting range to \autoref{subsec:robustness}.

The right panel of \autoref{fig:wtheta_cov} shows the corresponding jackknife \emph{correlation} matrix. We find non-negligible bin-to-bin correlations, particularly on larger angular scales where the same large-scale modes contribute to multiple \(\theta\) bins. This motivates the use of the full covariance matrix (rather than diagonal errors) in all subsequent power-law and \(\Lambda\)CDM template fits presented in later sections.

\subsection{Robustness to selection and masking choices}
\label{subsec:robustness}

A key requirement for using radio catalogues as large-scale-structure tracers is that the measured \(w(\theta)\) is not driven by residual survey systematics (e.g. depth gradients, edge effects, or source-finding artefacts). We therefore test the stability of the angular correlation function against a set of deliberately conservative variations of the fiducial selection and mask. These tests probe three main factors: (i) the adopted depth threshold used to define the footprint, (ii) the flux-density cut (and hence the source-population mix), and (iii) the treatment of complex/multi-component radio morphology. All measurements shown below adopt the fiducial RA-edge selection (\(\mathrm{RA}\le 184.07^\circ\)) unless stated otherwise.

\begin{figure*}
    \centering
    \includegraphics[width=\textwidth, trim=0 0 0 0, clip]{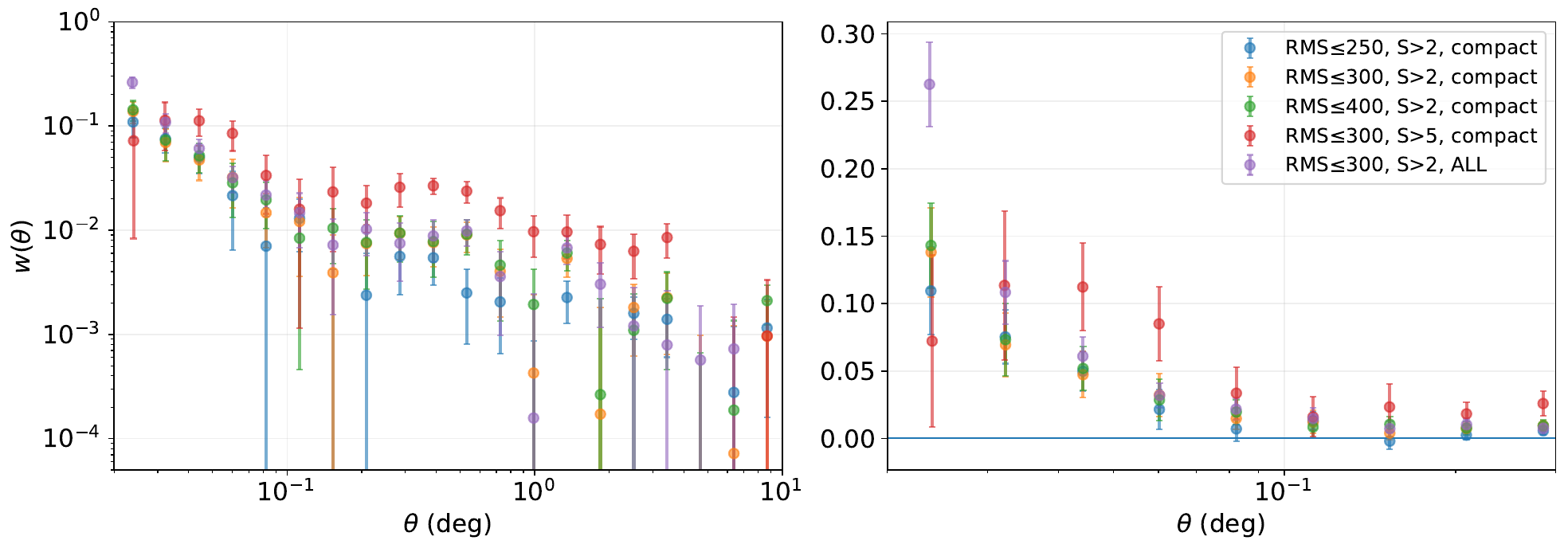}
    \caption{Stability of the measured $w(\theta)$ to alternative sample selections and masks. Curves show $w(\theta)$ for the fiducial case and for a selection of depth/flux/morphology variants. \textbf{Left:} \(w(\theta)\) over the full angular range. \textbf{Right:} a zoom of the small-angle regime highlighting morphology-driven effects. The compact-only samples show consistent clustering amplitudes on intermediate and large scales, indicating that the fiducial result is not driven by the precise depth cut. In contrast, the all-sources selection exhibits a pronounced small-scale excess, consistent with multi-component radio galaxies being resolved into multiple catalogue entries; this motivates our compact-only fiducial choice and the conservative fitting window adopted for parameter inference.}
    \label{fig:wtheta_stability}
\end{figure*}

\autoref{fig:wtheta_stability} compares the fiducial \(w(\theta)\) to the following variants:
\begin{itemize}
    \item \textbf{Depth-mask variations (compact, \(S_{816}\ge 2~\mathrm{mJy}\)):} we relax the rms threshold from the fiducial \(\sigma_{\rm local}\le 250~\mu\mathrm{Jy\,beam^{-1}}\) to \(\sigma_{\rm local}\le 300\) and \(400~\mu\mathrm{Jy\,beam^{-1}}\), thereby including progressively noisier regions closer to the footprint boundary.
    \item \textbf{Brighter flux cut (compact, \(\sigma_{\rm local}\le 300~\mu\mathrm{Jy\,beam^{-1}}\)):} we raise the flux threshold to \(S_{816}\ge 5~\mathrm{mJy}\) as a population-driven stress test.
    \item \textbf{All-sources morphology test (\(S_{816}\ge 2~\mathrm{mJy}\), \(\sigma_{\rm local}\le 300~\mu\mathrm{Jy\,beam^{-1}}\)):} we repeat the measurement without the compact-only restriction, i.e. including sources with multiple fitted Gaussians/components.
\end{itemize}

Across the full angular range, the compact-only samples selected with \(\sigma_{\rm local}\le 250\), 300, and \(400~\mu\mathrm{Jy\,beam^{-1}}\) yield consistent \(w(\theta)\) within the jackknife uncertainties (\autoref{fig:wtheta_stability}, left). This indicates that the fiducial clustering signal is not being driven by the noisiest edge regions that are preferentially added when the depth cut is relaxed. The agreement also supports our use of a binary, rms-threshold mask for defining the angular selection function.

Raising the flux threshold reduces the source surface density and thus increases the statistical uncertainties, but the overall shape of \(w(\theta)\) remains similar on intermediate and large scales. The \(S_{816}\ge 5\) mJy curve lies systematically above the \(S_{816}\ge 2\) mJy curves. This is expected because brighter radio samples are increasingly dominated by radio-loud AGN, which preferentially inhabit more massive dark-matter haloes than star-forming galaxies and therefore trace the matter field with a higher large-scale bias \citep{Hale_2018,Magliocchetti_2017}. We treat this as a qualitative consistency check rather than a primary measurement, since the reduced number of sources leads to larger errors.

The most significant deviation arises when the compactness requirement is removed. The all-sources selection exhibits a pronounced upturn in \(w(\theta)\) at small separations (\autoref{fig:wtheta_stability}, right), while remaining broadly consistent with the compact-only samples on larger scales. This behaviour is expected: although we use the SRL catalogue (which merges multiple Gaussians within a single island into one source entry), extended radio galaxies and complex emission can be fragmented into multiple islands and hence multiple SRL entries (and bright-source artefacts can add nearby spurious detections). These effects generate an artificial excess of close pairs, motivating either explicit component-association (‘collapsing’) or a conservative compact-only sample for interpreting small-scale clustering.

The angular scale where this effect becomes important provides a practical guide for defining a conservative minimum scale for modelling. For DR1 the typical restoring beam is \(\sim 32\arcsec \times 17\arcsec\), so even \(\theta=0.02^\circ\) corresponds to \(\sim 72\arcsec\) (a few beamwidths). While we present \(w(\theta)\) down to \(0.02^\circ\), \autoref{fig:wtheta_stability} shows that the all-sources curve departs strongly from the compact-only measurement below \(\theta \simeq 0.1^\circ\). To minimise sensitivity to source association choices and to avoid residual blending/deblending effects, all quantitative fits and bias inference in this work are therefore restricted to \(\theta \ge 0.112^\circ\).

\section{Phenomenological power-law fits}
\label{sec:powerlaw}

Before interpreting the clustering in terms of physical dark-matter models, we first characterise the measured signal using a standard phenomenological power-law form (\autoref{w_theta_eq}). We fit the power-law model using generalised least squares with the full jackknife covariance (\autoref{subsec:cov}); for fixed \(\gamma\), the best-fitting amplitude \(A\) is obtained analytically. We also report the corresponding goodness-of-fit using the reduced chi-squared, \(\chi^2_\nu\), with \(\nu = N_{\rm fit}-1\) degrees of freedom, where \(N_{\rm fit}\) is the number of fitted angular bins and the subtraction of 1 reflects the single free parameter \(A\) in the fixed-\(\gamma\) fit.

To minimise sensitivity to morphology-driven small-scale excess power and to avoid potential large-scale systematics, we restrict our primary fits to a conservative angular window of \(0.112^\circ \le \theta \le 1.36^\circ\) (spanning 9 logarithmic bins). This lower bound is motivated empirically by the morphology stress test in \autoref{fig:wtheta_stability}. The upper bound is chosen to remain in a regime where the measurement is not dominated by the largest-scale gradients and where the integral-constraint correction remains small (\autoref{subsec:ic}).

\begin{figure*}
    \centering
    \includegraphics[width=\textwidth, trim=0 0 0 0, clip]{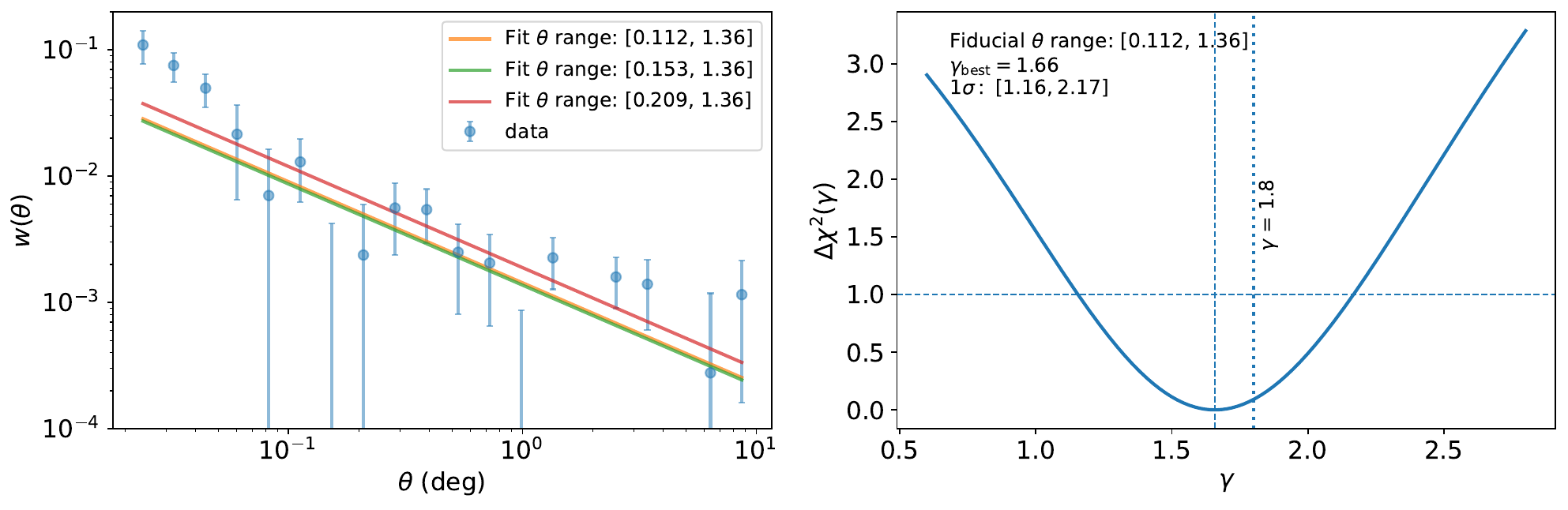}
    \caption{Power-law characterisation of the clustering signal. \textbf{Left:} fiducial \(w(\theta)\) measurement with jackknife errors together with best-fitting fixed-slope power-law models, \(w(\theta)=A(\theta/1^\circ)^{1-\gamma}\), with \(\gamma=1.8\), fitted over three angular windows that progressively exclude the smallest-separation bins: \([0.112,1.36]^\circ\) (fiducial), \([0.153,1.36]^\circ\), and \([0.209,1.36]^\circ\). The corresponding amplitudes are \(A(1^\circ)=(1.434\pm0.475)\times10^{-3}\), \((1.374\pm0.483)\times10^{-3}\), and \((1.890\pm0.672)\times10^{-3}\), with reduced \(\chi^2_\nu=0.63\), 0.74, and 0.75, respectively. \textbf{Right:} profile likelihood \(\Delta\chi^2(\gamma)\) for the slope parameter in the fiducial window, obtained by minimising over \(A\) at each fixed \(\gamma\); the dashed horizontal line marks \(\Delta\chi^2=1\) (1\(\sigma\) for one parameter), the dashed vertical line indicates the best-fitting \(\gamma\), and the dotted vertical line marks the canonical choice \(\gamma=1.8\) adopted for our baseline amplitude comparison.}
    \label{fig:wtheta_fit}
\end{figure*}

\autoref{fig:wtheta_fit} summarises the power-law characterisation. Following common practice in radio-continuum clustering analyses, where intermediate-scale \(w(\theta)\) is often well described by a power law with \(\gamma \simeq 1.7\)--1.9, we adopt the canonical fixed-slope choice \(\gamma=1.8\) to provide a stable reference amplitude and to enable straightforward comparison with the literature. In the left panel, we fit the amplitude over three angular windows that progressively exclude the smallest-separation bins: \([0.112,1.36]^\circ\) (fiducial), \([0.153,1.36]^\circ\), and \([0.209,1.36]^\circ\). The corresponding amplitudes are
\begin{align}
A(1^\circ) &= (1.434\pm0.475)\times 10^{-3}, \qquad (\chi^2_\nu=0.63), \nonumber\\
A(1^\circ) &= (1.374\pm0.483)\times 10^{-3}, \qquad (\chi^2_\nu=0.74), \nonumber\\
A(1^\circ) &= (1.890\pm0.672)\times 10^{-3}, \qquad (\chi^2_\nu=0.75). \nonumber
\end{align}
These values are consistent with each other within the statistical uncertainties\footnote{The sub-unity reduced \(\chi^2\) values should not be over-interpreted: with only \(\nu=6\)–8 degrees of freedom, substantial scatter about unity is expected, and the Hartlap scaling applied to the precision matrix further lowers the nominal \(\chi^2\) values. In this sense, they are best viewed as approximate goodness-of-fit diagnostics.}, indicating that the inferred clustering amplitude is not driven by a single bin at the lower edge of the fitting range. All three fits provide an acceptable description of the data on these intermediate scales.

In the right panel of \autoref{fig:wtheta_fit}, we assess how strongly the data constrain the slope \(\gamma\) by profiling the likelihood. For the slope constraints, we compute a profile likelihood \(\Delta\chi^2(\gamma)=\chi^2(\gamma)-\chi^2_{\min}\) by refitting \(A\) at each fixed \(\gamma\) using the full covariance. The best-fitting slope is \(\gamma_{\rm best}\simeq 1.66\), but the profile is broad, indicating that the present data do not tightly constrain \(\gamma\) over the adopted fitting range. In particular, the canonical choice \(\gamma=1.8\) is fully consistent with the data, and we therefore adopt it as a stable reference value for quoting \(A(1^\circ)\).

For the fiducial fixed-slope fit, \(A(1^\circ)=(1.434\pm0.475)\times10^{-3}\), corresponding to
\(\log_{10}A = -2.843^{+0.124}_{-0.175}\). This phenomenological description serves as a compact summary of the DR1 clustering signal and provides a useful reference point for the \(\Lambda\)CDM template-based bias interpretation presented in \autoref{sec:bias}.

\section{Bias inference from $\Lambda$CDM template modelling}
\label{sec:bias}

\subsection{Overview of bias interpretation}
\label{subsec:bias_overview}

The angular correlation function \(w(\theta)\) measured from a flux-limited radio catalogue is a projected statistic: it mixes contributions from different redshifts and physical scales because sources at different distances project the same angular separation \(\theta\) to different comoving separations. To connect the observed clustering to the underlying matter distribution, we compare our measurements to a \(\Lambda\)CDM prediction for the projected dark-matter angular correlation function, \(w_{\rm DM}(\theta)\), computed for an assumed source redshift distribution \(N(z)\). Details of the template construction and the adopted \(N(z)\) priors are given in \autoref{subsec:template_priors}.

On sufficiently large scales, where density fluctuations are in the linear regime, the clustering of a tracer population can be related to the matter field by a scale-independent linear bias factor \citep{Kaiser_1984,Mo_White_1996}. In angular space this implies that the observed correlation function can be written as an amplitude rescaling of the matter template,
\begin{equation}
w(\theta) \;\approx\; b_{\rm eff}^2\, w_{\rm DM}(\theta),
\label{eq:bias_basic}
\end{equation}
where \(b_{\rm eff}\) is an \emph{effective} bias that combines the redshift- and luminosity-dependent mix of the flux-limited sample into a single parameter. Because the projection involves pairs, \(b_{\rm eff}\) should be interpreted as a weighted average over the underlying \(b(z)\) of the contributing populations, with the weighting set primarily by \(N^2(z)\) and the projection kernel. In this work, we apply this linear, scale-independent bias model over the same conservative angular range used for the power-law fits.

As discussed in \autoref{subsec:ic}, the finite survey footprint suppresses the measured correlation function by a nearly constant offset (the integral constraint, IC) that depends on both the survey mask and the assumed model. For any proposed \(w_{\rm DM}(\theta)\), we therefore evaluate \({\rm IC}\) using the random--random weights (\autoref{eq:ic_rr}) and compare the data to
\begin{equation}
w_{\rm pred}(\theta) \;=\; b_{\rm eff}^2\left[w_{\rm DM}(\theta)-{\rm IC}\right].
\label{eq:bias_ic}
\end{equation}

Because \(w_{\rm DM}(\theta)\) is a projected template, the dominant modelling uncertainty is typically the redshift distribution \(N(z)\) for radio-continuum samples without individual redshifts. We therefore bracket this uncertainty by repeating the template calculation and bias fit for two plausible \(N(z)\) priors (AGN-dominated and TOTAL; discussed below). 

\subsection{Matter template and redshift-distribution priors}
\label{subsec:template_priors}

Interpreting \(w(\theta)\) in terms of bias requires a prediction for the projected matter correlation function \(w_{\rm DM}(\theta)\) for unbiased tracers (\(b=1\)). We compute the non-linear matter power spectrum with \textsc{CAMB} \citep{Lewis_2000, Lewis2000CAMB}, project it to an angular power spectrum using the Limber approximation \citep{Limber1953,LoVerdeAfshordi2008}. For an unbiased tracer, the projected matter angular power spectrum is calculated under the Limber approximation as
\begin{equation}
C_\ell^{\rm DM} = \int dz \, \frac{H(z)}{c}\,
\frac{N^2(z)}{\chi^2(z)}\,
P_{\rm m}\!\left(k=\frac{\ell+1/2}{\chi(z)}, z\right),
\end{equation}
where $\ell$ is the angular multipole, $H(z)$ is the Hubble parameter at redshift $z$, $c$ is the speed of light, $P_{\rm m}(k,z)$ is the non-linear matter power spectrum, $\chi(z)$ is the comoving radial distance, and $N(z)$ is the source redshift distribution. We then obtain the configuration-space matter template via
\begin{equation}
w_{\rm DM}(\theta)=\sum_{\ell}\frac{2\ell+1}{4\pi}C_\ell^{\rm DM}P_\ell(\cos\theta),
\end{equation}
which is used in the bias fit below.

The projection depends sensitively on the source redshift distribution \(N(z)\). Unlike optical samples, wide-area radio continuum catalogues generally lack redshifts for the bulk of sources, so \(N(z)\) must be supplied externally (e.g. from cross-matched photo-\(z\) catalogues or simulations). In this work we adopt semi-empirical \(N(z)\) priors derived from the T-RECS radio simulations \citep{Bonaldi_2019}, constructed to match a wide-area selection at 816\,MHz with \(S_{816}\ge 2~\mathrm{mJy}\). We use two bracketing priors to span plausible \(N(z)\) uncertainty for a bright mJy sample: (i) an AGN-only prior, representing an AGN-dominated limiting case rather than a literal assumption that no SFGs are present, and (ii) a TOTAL prior that includes both AGN and star-forming galaxies. The true redshift distribution is expected to lie between these two extremes. Both priors are normalised to unit area, \(\int N(z)\,dz = 1\), so that differences reflect the relative redshift weighting rather than an overall normalisation.

\begin{figure*}
    \centering
    \includegraphics[width=\textwidth, trim=0 0 0 0, clip]{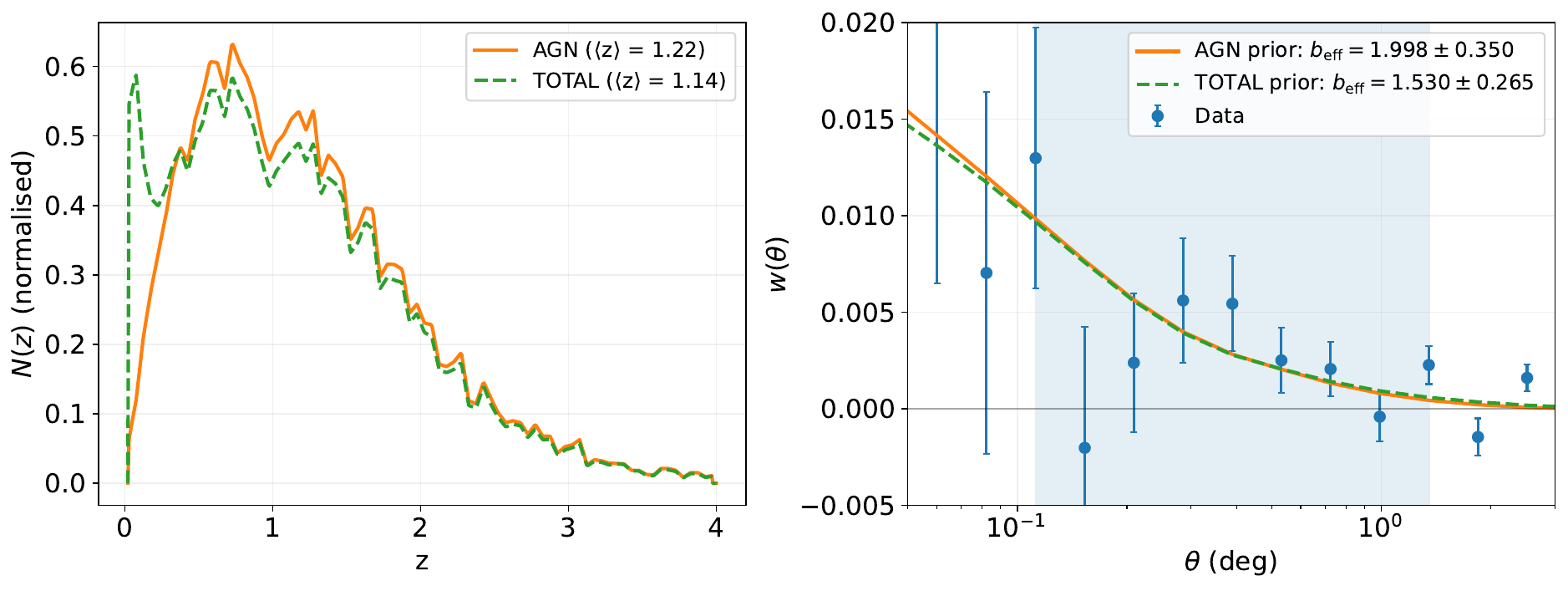}
    \caption{Bias inference from \(\Lambda\)CDM template modelling. \textbf{Left:} redshift-distribution priors \(N(z)\) for the \(S_{816}\ge2~\mathrm{mJy}\) sample, taken from the T-RECS semi-empirical simulations for an AGN-dominated selection and for the full population (TOTAL). Each \(N(z)\) is normalised to unit area; the mean redshift \(\langle z\rangle\) of each prior is indicated in the legend. \textbf{Right:} fiducial \(w(\theta)\) measurement with jackknife errors compared to the best-fitting projected matter templates for each \(N(z)\) prior. The shaded region marks the angular fitting window used for the bias inference.}
    \label{fig:bias_fit}
\end{figure*}

The adopted priors are shown in \autoref{fig:bias_fit} (left). The AGN prior peaks at \(z\simeq 0.73\) with mean \(\langle z\rangle \simeq 1.22\), while the TOTAL prior exhibits an additional low-redshift contribution (peak \(z\simeq 0.076\)) but retains a substantial high-\(z\) tail, yielding a similar mean \(\langle z\rangle \simeq 1.14\). The low-\(z\) peak but relatively high mean in the TOTAL case reflects this mixed population: a significant nearby star-forming component combined with a broad AGN tail to \(z\gtrsim 2\). By repeating the template calculation and bias inference for both priors we explicitly bracket the systematic uncertainty associated with \(N(z)\), which is typically the dominant limitation in radio-continuum clustering analyses.

\subsection{Fitting procedure and results}
\label{subsec:bias_results}

We infer an effective large-scale bias by fitting the measured \(w(\theta)\) with a \(\Lambda\)CDM projected-matter template. For a given redshift-distribution prior \(N(z)\) (\autoref{subsec:template_priors}), we compute the corresponding matter template \(w_{\rm DM}(\theta)\) and evaluate the integral-constraint correction for that template using the random--random weights (\autoref{eq:ic_rr}). We then fit a single amplitude parameter \(\alpha=b_{\rm eff}^2\) to
\begin{equation}
w_{\rm data}(\theta) \;=\; \alpha\left[w_{\rm DM}(\theta)-{\rm IC}\right],
\label{eq:bias_fit_alpha}
\end{equation}
using generalised least squares with the full jackknife covariance matrix (\autoref{subsec:cov}). The fit is restricted to the conservative angular window \(0.112^\circ \le \theta \le 1.36^\circ\) (shaded region in \autoref{fig:bias_fit}, right), ensuring that the inferred \(b_{\rm eff}\) is driven by the two-halo/linear-regime signal and is not contaminated by morphology-driven small-scale excess (\autoref{subsec:robustness}). The IC is recomputed self-consistently for each \(N(z)\) prior because it depends on the template and the survey mask.

For the AGN \(N(z)\) prior, we obtain
\begin{equation}
b_{\rm eff} = 1.998 \pm 0.350 \qquad (\chi^2_\nu \simeq 0.76),
\end{equation}
while for the TOTAL prior we find
\begin{equation}
b_{\rm eff} = 1.530 \pm 0.265 \qquad (\chi^2_\nu \simeq 0.73).
\end{equation}
The best-fitting templates are shown in \autoref{fig:bias_fit} (right). Both priors provide statistically acceptable fits over the adopted angular range, indicating that the measured signal is well described as a scaled projection of \(\Lambda\)CDM matter clustering once the IC correction is included.

The shift between the two inferred biases, \(\Delta b_{\rm eff}\simeq 0.47\), quantifies the dominant systematic uncertainty in this analysis: the unknown true \(N(z)\) (and hence population mix) of the flux-limited radio sample. The TOTAL prior includes additional low-redshift weight relative to the AGN prior (\autoref{fig:bias_fit}, left), which increases the projected matter correlation \(w_{\rm DM}(\theta)\) at fixed \(\theta\) because a given angular separation corresponds to smaller physical separations at lower redshift where the matter clustering amplitude is larger. Consequently, matching a fixed observed \(w(\theta)\) requires a smaller amplitude rescaling, yielding a lower \(b_{\rm eff}\) for the TOTAL prior than for the AGN prior. We therefore quote a conservative bracket \(b_{\rm eff}\simeq 1.5\)--2.0 for the DR1 clustering sample, with the spread dominated by \(N(z)\) rather than by statistical errors in \(w(\theta)\).

Interpreted qualitatively within standard halo-bias relations, \(b_{\rm eff}\sim 2\) is consistent with radio sources residing predominantly in group-scale dark-matter haloes at \(z\sim 1\). However, because \(b_{\rm eff}\) is a population-weighted effective bias and the mapping to a characteristic halo mass depends on assumptions about halo occupation and source classification, we do not quote a single representative halo mass in this work.

\subsection{Spatial correlation length from Limber inversion}
\label{subsec:r0_limber}

To complement the bias constraints from template modelling, we also compute the real-space correlation length $r_0$, which encapsulates the clustering amplitude as a characteristic comoving scale. We derive $r_0$ by Limber-inverting the best-fitting power-law model for $w(\theta)$, using the same AGN and TOTAL $N(z)$ priors adopted in our analysis; this makes explicit how the inferred spatial clustering depends on the assumed redshift distribution. The resulting $r_0$ values should be interpreted as a derived, model-dependent summary rather than an independent measurement, but they provide a convenient point of comparison with other radio clustering studies.

We begin by adopting the standard power-law form for the real-space two-point correlation function,
\begin{equation}
\xi(r) = \left(\frac{r}{r_0}\right)^{-\gamma},
\label{eq:xi_powerlaw}
\end{equation}
and the corresponding angular form fitted in \autoref{w_theta_eq}. Under the (small-angle) Limber approximation, projecting $\xi(r)$ along the line of sight yields a power law in angle with amplitude related to $r_0$ by
\begin{equation}
w(\theta) \;=\; H_\gamma\, r_0^\gamma \,\theta^{1-\gamma}\,
\frac{\int dz\, \frac{H(z)}{c}\, N^2(z)\, \chi^{\,1-\gamma}(z)}
{\left[\int dz\, N(z)\right]^2}.
\label{eq:limber_wtheta}
\end{equation}
The geometrical prefactor is
\begin{equation}
H_\gamma \;=\; \sqrt{\pi}\;
\frac{\Gamma\!\left(\frac{\gamma-1}{2}\right)}{\Gamma\!\left(\frac{\gamma}{2}\right)}.
\label{eq:Hgamma}
\end{equation}
In practice, we use the same $N(z)$ priors shown in \autoref{fig:bias_fit} (left). Defining the Limber kernel integral
\begin{equation}
I_\gamma \;\equiv\; \frac{\int dz\, \frac{H(z)}{c}\, N^2(z)\, \chi^{\,1-\gamma}(z)}
{\left[\int dz\, N(z)\right]^2},
\label{eq:Igamma}
\end{equation}
the correlation length follows directly as
\begin{equation}
r_0 \;=\; \left[\frac{A}{H_\gamma\, I_\gamma}\right]^{1/\gamma}.
\label{eq:r0_from_A}
\end{equation}
When $\gamma$ is fixed, the uncertainty in $r_0$ from the fitted amplitude alone is
\begin{equation}
\frac{\sigma_{r_0}}{r_0} \;=\; \frac{1}{\gamma}\,\frac{\sigma_A}{A},
\label{eq:r0_error}
\end{equation}
where $\sigma_A$ is the (jackknife) uncertainty on $A$ from the power-law fit. This propagation does not include additional uncertainty from $N(z)$; in this work we instead bracket that dominant systematic by repeating the calculation for the AGN and TOTAL $N(z)$ priors.

It is also useful to characterise the redshift at which the Limber inversion is effectively weighted. We define a pair-weighted effective redshift
\begin{equation}
z_{\rm eff} \;=\; \frac{\int dz\, z\, W_\gamma(z)}{\int dz\, W_\gamma(z)},
\qquad
W_\gamma(z) \equiv \frac{H(z)}{c}\,N^2(z)\,\chi^{\,1-\gamma}(z),
\label{eq:zeff_def}
\end{equation}
and quote the mean $\langle z\rangle$ of the adopted $N(z)$ for reference.

Using our fiducial fixed-slope fit with $\gamma=1.8$ over $0.112^\circ \le \theta \le 1.36^\circ$ and the $N(z)$ priors in \autoref{subsec:template_priors}, we obtain:
$r_0=6.18\pm1.13\,h^{-1}$ Mpc (AGN) and $r_0=5.59\pm1.02\,h^{-1}$ Mpc (TOTAL). These values are consistent with the range typically reported for $\sim$GHz, mJy-selected radio samples when analysed with similar Limber-inversion assumptions, and provide a complementary summary statistic to our primary bias constraints from full $\Lambda$CDM template modelling (\autoref{subsec:bias_results}).

For comparison with studies that convert a power-law correlation length into a linear bias, we also translate our Limber-inverted \(r_0\) values into \(b(z)\) by relating the variance of a power-law \(\xi(r)=(r/r_0)^{-\gamma}\) in spheres of radius \(R\) to the matter variance, \(\sigma_{\rm m}(R,z)=\sigma_8\,D(z)/D(0)\) \citep[e.g.][]{Peebles1980}. Here \(D(z)\) is the linear growth factor of matter fluctuations, normalised such that \(D(0)=1\), so that the matter fluctuation amplitude evolves as \(\sigma_{\rm m}(R,z)=\sigma_{\rm m}(R,0)\,D(z)\). For a top-hat filter this gives \(\sigma_{\rm g}^2(R)=J_2(\gamma)\,(r_0/R)^\gamma\), with \(J_2(\gamma)=72/[(3-\gamma)(4-\gamma)(6-\gamma)2^\gamma]\), and hence
\begin{equation}
b(z) \;=\; \left(\frac{r_0}{8\,h^{-1}{\rm Mpc}}\right)^{\gamma/2}
\frac{\sqrt{J_2(\gamma)}}{\sigma_8\,D(z)/D(0)}.
\end{equation}
Evaluating this expression at the pair-weighted effective redshifts \(z_{\rm eff}\) from \autoref{eq:zeff_def} for our fiducial cosmology), we obtain
\(b(z_{\rm eff})=2.18\pm0.36\) for the AGN prior (\(z_{\rm eff}=0.984\)) and
\(b(z_{\rm eff})=1.78\pm0.29\) for the TOTAL prior (\(z_{\rm eff}=0.740\)).
These values are consistent in scale with our primary template-based \(b_{\rm eff}\) constraints (\autoref{subsec:bias_results}).

\section{Comparison with previous radio angular clustering measurements}
\label{sec:comparison}

Angular clustering of radio-selected sources has now been measured across a wide range of frequencies, depths, and sky areas \citep{Magliocchetti_1999,Magliocchetti_2004,Negrello_2006,Lindsay2014a,Nusser_2015,Siewert2020,Tiwari_2022,Hale_2024}. Direct comparison between surveys is not strictly one-to-one because the inferred clustering amplitude and bias depend on several coupled factors: the adopted flux threshold (and hence the evolving mix of radio-loud AGN and star-forming galaxies; e.g. \citealt{Wilman_2008,Magliocchetti_2017,Hale_2018,Bonaldi_2019}), the treatment of resolved multi-component sources at small separations, the assumed redshift distribution $N(z)$ used to interpret the signal, and survey-specific systematics such as spatially varying depth, smearing, or calibration artefacts. With those caveats in mind, MeerKLASS UHF DR1 provides a useful new reference point at 816\,MHz, bridging the gap between the classic $\sim$GHz radio surveys and the new generation of wide-area low-frequency surveys.

A common phenomenological summary of radio clustering is the power-law amplitude at a pivot angle of $1^\circ$ (\autoref{w_theta_eq}). For our fiducial compact, depth-controlled DR1 sample we obtain
$A(1^\circ)=(1.434\pm0.475)\times10^{-3}$ for fixed $\gamma=1.8$, corresponding to
$\log_{10}A=-2.843^{+0.124}_{-0.175}$.
This lies comfortably within the broad envelope of published radio measurements at mJy-like thresholds \citep{deZotti_2010,Magliocchetti_2022}.

Early wide-area analyses at $\sim$GHz frequencies established two key empirical features: (i) a steep small-angle excess caused by multi-component radio galaxies, and (ii) a shallower cosmological clustering term on larger scales that is well described by a near-canonical slope, $\gamma\simeq1.8$, once multi-component effects are controlled \citep{Blake_Wall_2002,BlakeWall2002}. Similar conclusions were reached for other wide-area catalogues such as SUMSS and WENSS, with large-scale amplitudes of order $A\sim10^{-3}$ when expressed in comparable parameterisations \citep{BlakeMauchSadler2004}. More recent interferometric surveys have extended clustering measurements to fainter limits and different frequencies, often quoting $\log_{10}A$ for ease of comparison across orders of magnitude. Representative examples include FIRST at 1.4\,GHz, TGSS-ADR at 150\,MHz, deep-field measurements such as ELAIS-N1 at 400\,MHz, LoTSS-DR1 at 144\,MHz, and more recently LoTSS-DR2 \citep{Lindsay2014a,Lindsay2014b,RanaBagla2019,Chakraborty_2020,Siewert2020,Hale_2024,Mazumder_2022}. Within this landscape, our $\log_{10}A=-2.843^{+0.124}_{-0.175}$ is somewhat lower than the amplitudes inferred in some smaller-area fields, where sample variance and source-association choices can play a larger role \citep{Heywood_2013}, but remains fully consistent with the broad range of values reported for modern radio surveys.

Several methodological points are worth emphasizing when comparing amplitudes. First, the fitted value of $A$ depends modestly on the adopted angular fit window: the smallest-angle bins are most sensitive to residual component-splitting and association choices, whereas the largest angles are more susceptible to integral-constraint corrections and residual large-scale survey specific systematics. Our stability tests (\autoref{fig:wtheta_fit}) show that $A(1^\circ)$ remains consistent, within the uncertainties, when progressively excluding the smallest-separation bins. Second, flux thresholds defined at different observing frequencies are not directly interchangeable. A simple spectral-index conversion does not guarantee that two surveys select the same underlying luminosity, redshift, and AGN/SFG mixture, so cross-survey comparisons remain only approximate even when the nominal flux cuts appear similar. This is likely one reason for the intrinsic spread in reported $\log_{10}A$ values. Consistent with this interpretation, our own brighter subsample ($S_{816}\geq5$\,mJy) lies systematically above the fiducial $S_{816}\geq2$\,mJy sample on intermediate and large scales (\autoref{fig:wtheta_stability}), as expected if brighter cuts preferentially select more strongly biased AGN-dominated populations. Third, literature measurements are not always derived with identical covariance treatment; some earlier studies used diagonal-only errors \citep{Lindsay2014a}, whereas our analysis uses the full spatial jackknife covariance throughout. This should be borne in mind when comparing absolute best-fitting amplitudes and biases.

A comparison of effective bias measurements is more delicate still, because the inferred bias depends not only on flux limit and observing frequency, but also on the assumed $N(z)$, the adopted bias parametrisation, and whether the sample is treated as a single effective tracer or decomposed into AGN and SFG sub-populations. Bright radio samples are typically AGN-dominated, while fainter samples contain an increasing contribution from star-forming galaxies; since AGN preferentially inhabit more massive haloes, they are expected to exhibit a larger large-scale bias than SFGs \citep{Magliocchetti_2017,Hale_2018,Chakraborty_2020,Mazumder_2022}. Our DR1 sample is not population-split, and is additionally restricted to compact single-component sources, so the inferred $b_{\rm eff}$ should be interpreted as a pair-weighted average over a mixed population rather than as the bias of a single physical class.

For MeerKLASS DR1, our template-based modelling yields
$b_{\rm eff}=1.998\pm0.350$ for an AGN-dominated $N(z)$ prior and
$b_{\rm eff}=1.530\pm0.265$ for a TOTAL prior (\autoref{subsec:bias_results}), explicitly bracketing the dominant modelling uncertainty associated with the redshift distribution. These values are broadly consistent with the long-standing picture that mJy-selected radio samples are dominated by radio-loud AGN hosted by group-scale haloes, leading to typical effective biases of order $b\sim1.5$--3 depending on flux limit, redshift, and population mix \citep{Mo_White_1996, Tinker_2010}. They are also comparable in scale to recent wide-area low-frequency results. In particular, the LoTSS-DR2 analysis found $b_{\rm C}=2.14^{+0.22}_{-0.20}$ for a constant-bias model and $b_{\rm E}(z=0)=1.79^{+0.15}_{-0.14}$ for an evolving model inversely proportional to the growth factor, corresponding to $b_{\rm E}\simeq2.81$ at the median redshift of that sample \citep{Hale_2024}. Given the different observing frequency, sky area, flux limit, source selection, and redshift calibration, the level of agreement is encouraging.

It is also instructive to compare our results with studies that jointly constrain the redshift distribution and bias evolution. Using LoTSS-DR1 together with external information, \citet{Alonso_2021} showed that the inferred bias depends sensitively on the assumed $N(z)$ and on whether one adopts a constant or evolving bias model. This is qualitatively consistent with our own result that the inferred $b_{\rm eff}$ shifts appreciably between the AGN-prior and TOTAL-prior cases. More generally, comparisons to studies that explicitly separate AGN and SFG populations should be made with care: agreement with AGN-only measurements is not necessarily expected for a mixed sample such as ours, for which a somewhat lower effective bias is natural.

For completeness, we also quote Limber-inverted correlation lengths for $\gamma=1.8$,
$r_0=6.18\pm1.13\,h^{-1}\mathrm{Mpc}$ for the AGN prior and
$r_0=5.59\pm1.02\,h^{-1}\mathrm{Mpc}$ for the TOTAL prior
(\autoref{subsec:r0_limber}). These values are in line with classic radio-clustering results such as NVSS, which inferred characteristic correlation lengths of a few to several $h^{-1}\,\mathrm{Mpc}$ under reasonable redshift-distribution assumptions \citep{BlakeWall2002}. However, such comparisons should be interpreted cautiously. Limber-inverted $r_0$ or bias values derived from an assumed power-law spatial correlation function are not strictly equivalent to the effective bias inferred from projected matter-template fits, and results obtained under comoving-clustering assumptions are more appropriately compared to evolving-bias rather than constant-bias models.

Taken together, our results reinforce a coherent picture emerging from previous radio surveys: once depth systematics and multi-component sources are carefully controlled, radio angular clustering on intermediate scales is well described by a modest-amplitude power law, while the corresponding effective bias is consistent with a predominantly AGN-driven population diluted by an increasing SFG contribution towards fainter flux densities.

\section{Conclusions}
\label{sec:conclusions}

We have presented the first measurement of the angular clustering of radio sources from the MeerKLASS UHF \emph{on-the-fly} continuum survey Data Release~1. The constant-elevation, cross-linked rising/setting scan strategy delivers large-area mapping efficiency together with a stable, well-characterised PSF and predictable overlap-driven depth patterns, enabling wide-area clustering analyses with an explicitly modelled selection function. Using the DR1 SRL catalogue at a reference frequency of 816\,MHz, we constructed a conservative clustering sample designed to minimise depth-driven systematics and morphology-induced small-scale excess. Our fiducial selection applies a flux-density threshold \(S_{816}\ge2~\mathrm{mJy}\), a depth-controlled footprint defined by \(\sigma_{\rm local}\le250~\mu\mathrm{Jy\,beam^{-1}}\) from the DR1 rms map, a compact-only criterion, and an empirically motivated RA-edge cut (\(\mathrm{RA}\le184.07^\circ\)) validated through data--random uniformity diagnostics. Random catalogues were generated using an RMS-based accept--reject procedure that reproduces the survey mask and selection function, and uncertainties were estimated from a spatial jackknife covariance.

Our main results are:
\begin{enumerate}
    \item We detect a positive clustering signal over a broad range of angular separations (\(0.02^\circ\lesssim\theta\lesssim10^\circ\)). The jackknife correlation matrix shows non-negligible bin-to-bin covariances, motivating the use of the full covariance in all model fits.
    \item Robustness tests demonstrate that the measured \(w(\theta)\) is stable to reasonable variations in the depth mask and flux threshold on intermediate and large scales. Relaxing compactness criteria produces a pronounced small-scale upturn, consistent with fragmentation of extended/multi-component radio galaxies into multiple catalogue entries, validating our compact-only fiducial choice and motivating a conservative minimum scale for interpretation.
    \item A phenomenological power-law description provides a compact summary of the intermediate-scale signal. Fixing \(\gamma=1.8\), we obtain a fiducial amplitude \(A(1^\circ)=(1.434\pm0.475)\times10^{-3}\) over \(0.112^\circ\le\theta\le1.36^\circ\), corresponding to \(\log_{10}A=-2.843^{+0.124}_{-0.175}\). Varying the lower bound of the fit window yields consistent amplitudes within uncertainties, indicating that the result is not driven by the smallest included bin.
    \item Interpreting the measurement in a \(\Lambda\)CDM framework, we fit a one-parameter amplitude \(\alpha=b_{\rm eff}^2\) to projected matter templates \(w_{\rm DM}(\theta)\) computed using \textsc{CAMB} and Limber projection, including an integral-constraint correction evaluated from random--random weights. The inferred effective bias depends primarily on the assumed redshift distribution \(N(z)\). Using two bracketing T-RECS priors, we obtain \(b_{\rm eff}=1.998\pm0.350\) (AGN prior) and \(b_{\rm eff}=1.530\pm0.265\) (TOTAL prior), quantifying the dominant modelling uncertainty from \(N(z)\).
    \item As an additional derived summary, we Limber-invert the power-law amplitude to obtain spatial correlation lengths \(r_0=6.18\pm1.13~h^{-1}\mathrm{Mpc}\) (AGN prior) and \(r_0=5.59\pm1.02~h^{-1}\mathrm{Mpc}\) (TOTAL prior) for \(\gamma=1.8\), and we report the corresponding effective redshifts for the Limber weighting. These values provide a convenient comparison point with previous radio clustering studies, subject to the usual power-law and \(N(z)\) assumptions.
\end{enumerate}

This work complements existing measurements by providing a wide-area (\(\sim800~\mathrm{deg}^2\)) clustering measurement at an intermediate GHz frequency. In this sense, DR1 occupies parameter space between classic all-sky GHz surveys (e.g. NVSS/SUMSS) and the new generation of low-frequency wide-area surveys (e.g. TGSS/LoTSS), while demonstrating that OTF interferometric continuum imaging can support precision large-scale-structure measurements. The overlap with contemporary optical spectroscopy in the DESI footprint motivates a clear next step: as spectroscopic and high-quality photometric redshifts become available for large fractions of the radio catalogue, future work can replace simulation-based \(N(z)\) priors with empirically calibrated redshift distributions and measure bias evolution and population-split clustering with substantially reduced modelling uncertainty.

\section*{Acknowledgments} 
SP acknowledges support from the Science and Technology Facilities Council (STFC) through the Consolidated Grant ST/X001229/1 at the Jodrell Bank Centre for Astrophysics, University of Manchester. SC acknowledges financial support from the South African National Research Foundation (Grant No. 84156) and the Inter-University Institute for Data Intensive Astronomy (IDIA). IDIA is a partnership of the University of Cape Town, the University of Pretoria and the University of the Western Cape. IDIA is registered on the Research Organization Registry with ROR ID 01edhwb26, and on Open Funder Registry with funder ID 100031500. SM and JM acknowledge the support provided by the German Federal Ministry of Education and Research (BMBF) through the BMBF D-MeerKAT III award (number 05A23WM2). OMS's research is supported by the South African Research Chairs Initiative of the Department of Science and Technology and National Research Foundation (grant No. 81737). The MeerKAT telescope is operated by the South African Radio Astronomy Observatory, which is a facility of the National Research Foundation, an agency of the Department of Science and Innovation. We acknowledge the use of the ilifu cloud computing facility – www.ilifu.ac.za, a partnership between the University of Cape Town, the University of the Western Cape, Stellenbosch University, Sol Plaatje University and the Cape Peninsula University of Technology. The ilifu facility is supported by contributions from the Inter-University Institute for Data Intensive Astronomy (IDIA – a partnership between the University of Cape Town, the University of Pretoria and the University of the Western Cape), the Computational Biology division at UCT and the Data Intensive Research Initiative of South Africa (DIRISA). This work made use of the CARTA (Cube Analysis and Rendering Tool for Astronomy) software (DOI: 10.5281/zenodo.3377984 – https://cartavis.github.io). We thank the developers of open-source Python libraries \textsc{NumPy} \citep{Numpy}, \textsc{SciPy} \citep{Scipy}, \textsc{Matplotlib} \citep{Matplotlib}, and \textsc{Astropy} \citep{astropy:2022}.

\section*{Data availability}
The data used in this study are available in the SARAO-hosted MeerKLASS data release at: \url{https://doi.org/10.48479/h9k3-7294}.

\bibliographystyle{mnras}
\bibliography{ref} 

\appendix

\section{Additional sample-selection diagnostics}
\label{app:selection_diagnostics}

\begin{figure*}
  \centering
  \includegraphics[width=\textwidth]{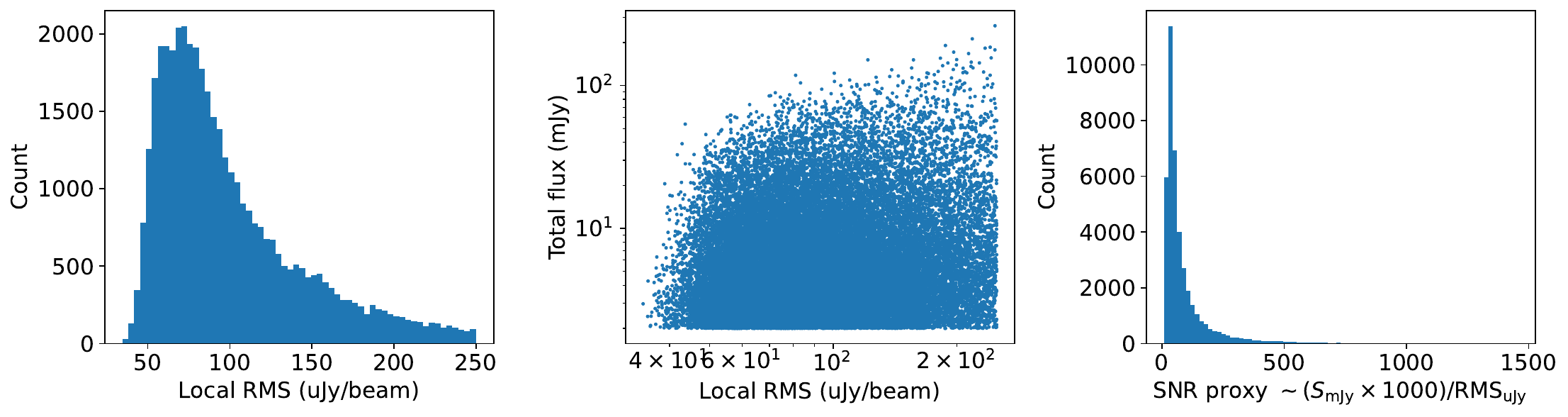}
  \caption{Depth/quality diagnostics for the fiducial clustering sample. \textbf{Left:} distribution of local rms noise \(\sigma_{\rm local}\) (from the DR1 rms map) evaluated at the selected source positions, showing that most sources lie well within the depth-controlled footprint and that the high-noise tail is suppressed by the \(\sigma_{\rm local}\le 250~\mu\mathrm{Jy\,beam^{-1}}\) cut. \textbf{Middle:} total integrated flux density versus \(\sigma_{\rm local}\), illustrating the combined action of the flux threshold \(S_{816}\ge 2~\mathrm{mJy}\) and the depth mask (vertical truncation at the rms boundary). \textbf{Right:} distribution of an approximate detection-significance proxy \(\mathrm{SNR}_{\rm proxy}\simeq (S_{\rm mJy}\times1000)/\sigma_{\mu\mathrm{Jy}}\), confirming that the fiducial flux cut corresponds to high-significance detections across the depth-controlled area.}
  \label{fig:app_A1_depth_quality}
\end{figure*}

\begin{figure*}
  \centering
  \includegraphics[width=\textwidth]{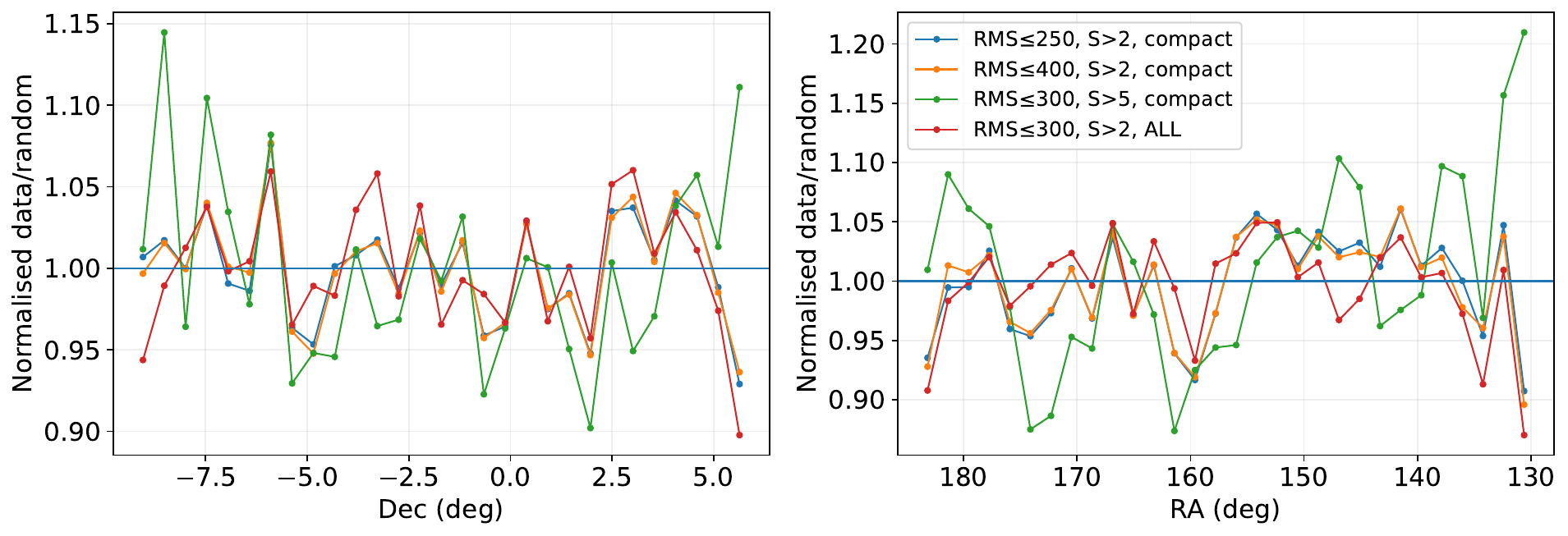}
  \caption{Selection-uniformity diagnostics for several alternative selections, shown as the ratio of normalised data to random counts \(R\equiv n_{\rm data}/n_{\rm rand}\) binned in declination (left; equal-width bins) and right ascension (right; equal-area bins defined by random-quantile boundaries). Error bars denote approximate \(1\sigma\) Poisson uncertainties propagated from the binned data and random counts. The profiles fluctuate around unity with no coherent large-scale trend, indicating that within the RA-restricted footprint the depth-controlled selection is spatially uniform and stable to reasonable changes in depth/flux/morphology cuts.}
  \label{fig:app_A2_uniformity_comparison}
\end{figure*}

The main analysis adopts a conservative clustering sample defined by a flux-density threshold, a depth-controlled footprint, and an empirically motivated RA-edge cut. This appendix provides supplementary diagnostics that guided these choices and helps illustrate why the adopted selection is well suited for large-scale clustering. Throughout, the data-to-random ratios are computed from counts normalised by the total number of objects in each catalogue (so that \(R\equiv n_{\rm data}/n_{\rm rand}=1\) is expected for a perfectly uniform selection within the mask); departures from unity therefore highlight spatial selection effects rather than changes in the overall surface density.

\autoref{fig:app_A1_depth_quality} summarises simple depth/quality diagnostics for the fiducial compact sample. The distribution of local rms values evaluated at the selected source positions (left panel) shows that the catalogue is predominantly drawn from well-behaved regions of the footprint, with the high-noise tail effectively suppressed by the adopted depth cut \(\sigma_{\rm local}\le 250~\mu\mathrm{Jy\,beam^{-1}}\). The middle panel shows total integrated flux density versus local rms, demonstrating that the flux cut and the depth mask act in complementary ways: the depth cut removes high-rms areas of the footprint (a vertical truncation in the plot), while the flux threshold removes faint sources (a horizontal truncation), so the final sample is not dominated by near-threshold detections in the noisiest regions. The right panel shows a simple signal-to-noise proxy, \(\mathrm{SNR}_{\rm proxy}\simeq (S_{\rm mJy}\times1000)/\sigma_{\mu\mathrm{Jy}}\), confirming that the \(S_{816}\ge2~\mathrm{mJy}\) selection lies safely above the effective detection limit for the depth-controlled footprint. In particular, at the fiducial depth boundary the proxy corresponds to \(\mathrm{SNR}_{\rm proxy}\gtrsim 8\), while the majority of sources have substantially higher values because they lie in regions with \(\sigma_{\rm local}\) well below the threshold.

\autoref{fig:app_A2_uniformity_comparison} compares the selection-uniformity diagnostic
\(R(\alpha,\delta)\equiv n_{\rm data}/n_{\rm rand}\) for several alternative selections used in \autoref{subsec:robustness}. The declination profiles fluctuate around unity with no coherent trend, indicating that the depth-controlled selection does not imprint a systematic gradient in \(\delta\). The right-ascension profiles are likewise broadly consistent with unity within the RA-restricted footprint, with modest bin-to-bin excursions that vary between selections (most noticeably for relaxed depth cuts that include noisier regions). This provides a useful sanity check that, once the RA-edge cut is enforced, reasonable changes in the depth/flux/morphology selection do not introduce large-scale density gradients that could mimic cosmological clustering.

\section{Robustness to angular fit window} \label{app:fit_window} 
\begin{figure}
  \centering
  \includegraphics[width=\columnwidth]{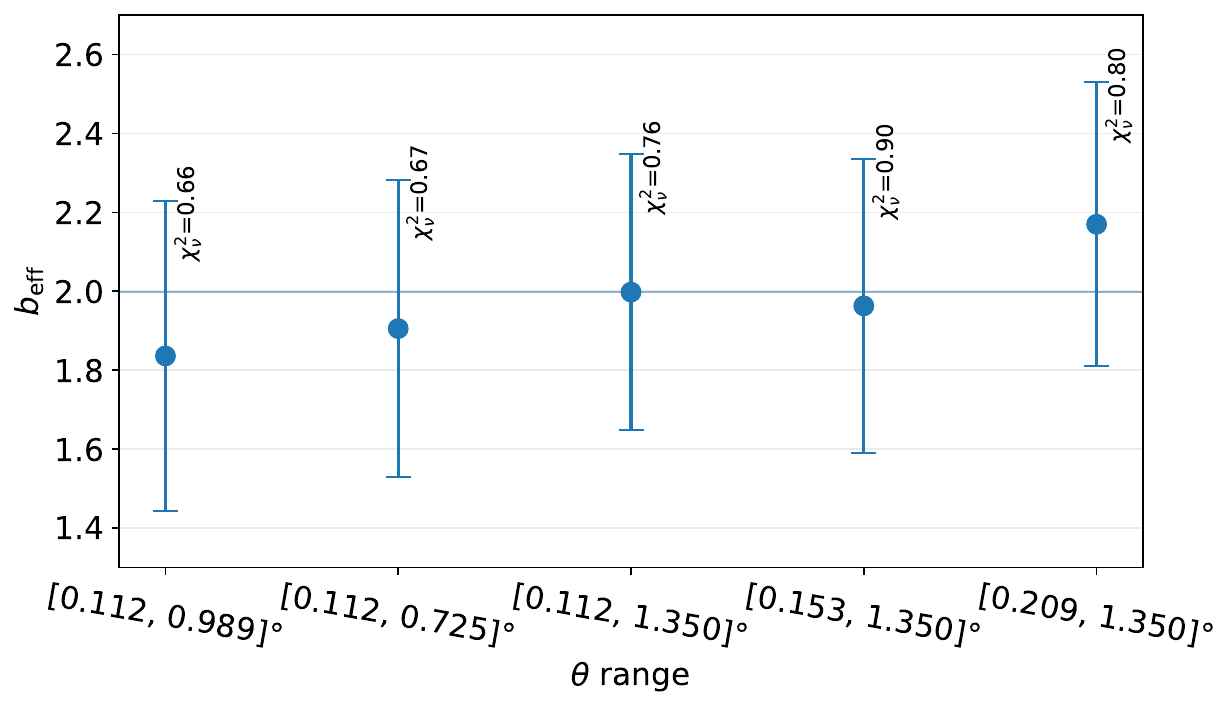}
  \caption{Stability of the inferred effective bias to the choice of angular fitting window. Each point shows the best-fitting \(b_{\rm eff}\) obtained by fitting the \(\Lambda\)CDM projected-matter template \(w_{\rm pred}(\theta)=b_{\rm eff}^2\,[w_{\rm DM}(\theta)-{\rm IC}]\) over the labelled angular range (lower and upper bounds in degrees), using the full covariance matrix. Error bars denote \(1\sigma\) uncertainties from the one-parameter fit, and the annotated values show the corresponding reduced chi-squared \(\chi^2_\nu\). The horizontal line indicates the fiducial best-fit value for the fiducial window \(0.112^\circ\le\theta\le1.36^\circ\). The inferred \(b_{\rm eff}\) is consistent within uncertainties across all tested windows, indicating that the bias constraint is not driven by a particular bin or by the precise choice of the fit range.}
  \label{fig:app_beff_stability}
\end{figure}

Our template-based bias inference (\autoref{subsec:bias_results}) assumes that the measured \(w(\theta)\) is well described, over a chosen angular range, by a scale-independent rescaling of the projected matter template. In practice, the inferred amplitude can depend on the adopted fit window: at the smallest separations residual morphology-driven effects or non-linear clustering may bias the fit, while at the largest separations the measurement becomes increasingly dominated by sample variance and is more sensitive to footprint-related effects (including the integral constraint). To test for any such dependence, we repeat the one-parameter template fit (\(\alpha=b_{\rm eff}^2\)) over a set of alternative angular windows that vary both the minimum and maximum scales. \autoref{fig:app_beff_stability} summarises the resulting best-fitting \(b_{\rm eff}\) values and their \(1\sigma\) uncertainties. The horizontal line shows the fiducial best-fit value for reference. Across all tested windows the inferred bias values are mutually consistent within the statistical uncertainties, and the reduced chi-squared values remain close to unity, indicating that no single angular bin or edge of the fitting range is driving the result. This stability supports the robustness of our fiducial choice \(0.112^\circ \le \theta \le 1.36^\circ\), which was adopted to avoid the morphology-driven small-scale excess while retaining adequate signal-to-noise for the template fit.

\section{Random catalogue convergence}
\label{app:random_convergence}
\begin{figure}
  \centering
  \includegraphics[width=\columnwidth]{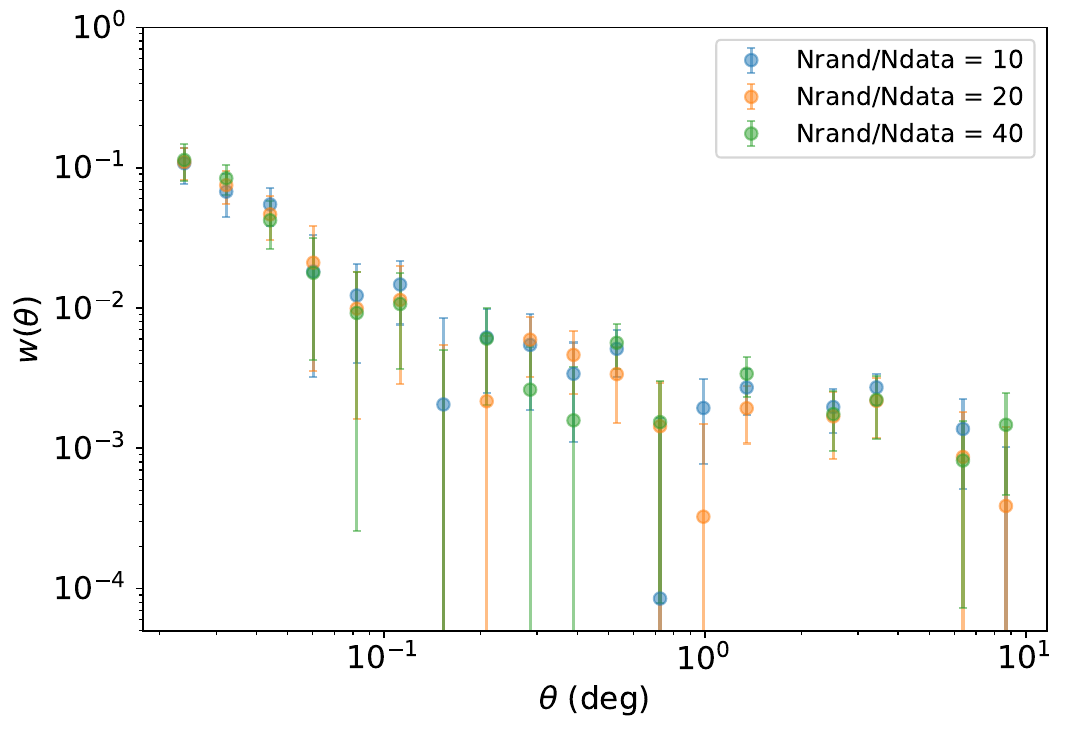}
  \caption{Convergence of the \(w(\theta)\) measurement with random-catalogue density. Points show \(w(\theta)\) measured with the Landy--Szalay estimator using random catalogues containing \(N_{\rm rand}/N_{\rm data}=10\), 20 (fiducial), and 40, with all other analysis choices held fixed. Error bars show the same jackknife uncertainties (from the same patching scheme) used in the main analysis. The curves are consistent within uncertainties, indicating that Poisson noise in the \(DR\) and \(RR\) pair counts is negligible for \(N_{\rm rand}/N_{\rm data}\ge 20\).}
  \label{fig:app_rand_cat_convergence}
\end{figure}

The Landy--Szalay estimator (\autoref{subsec:estimator}) uses a random catalogue to numerically encode the survey footprint and depth selection through the \(DR\) and \(RR\) pair counts. If the random catalogue is too sparse, shot noise in \(DR\) and \(RR\) can propagate into the estimator and appear as excess scatter in \(w(\theta)\), particularly at large separations where the number of pairs per bin is smaller. Our fiducial analysis adopts a random density \(N_{\rm rand}=20\,N_{\rm data}\); here we verify that this choice is safely in the converged regime.

\autoref{fig:app_rand_cat_convergence} compares \(w(\theta)\) measured using random catalogues with \(N_{\rm rand}/N_{\rm data}=10\), 20, and 40, keeping all other choices fixed (sample selection, mask, binning, and jackknife partition). The three measurements are statistically consistent across the full angular range, with no systematic shifts in amplitude and no visible reduction in scatter beyond the jackknife uncertainties when increasing \(N_{\rm rand}\). This confirms that random shot noise is sub-dominant for \(N_{\rm rand}/N_{\rm data}\ge 20\), and supports our fiducial choice as a computationally efficient setting that renders \(DR\) and \(RR\) noise negligible compared to the data and sample-variance uncertainties.

\section{Jackknife error robustness}
\label{app:jk_robustness}
\begin{figure}
  \centering
  \includegraphics[width=\columnwidth]{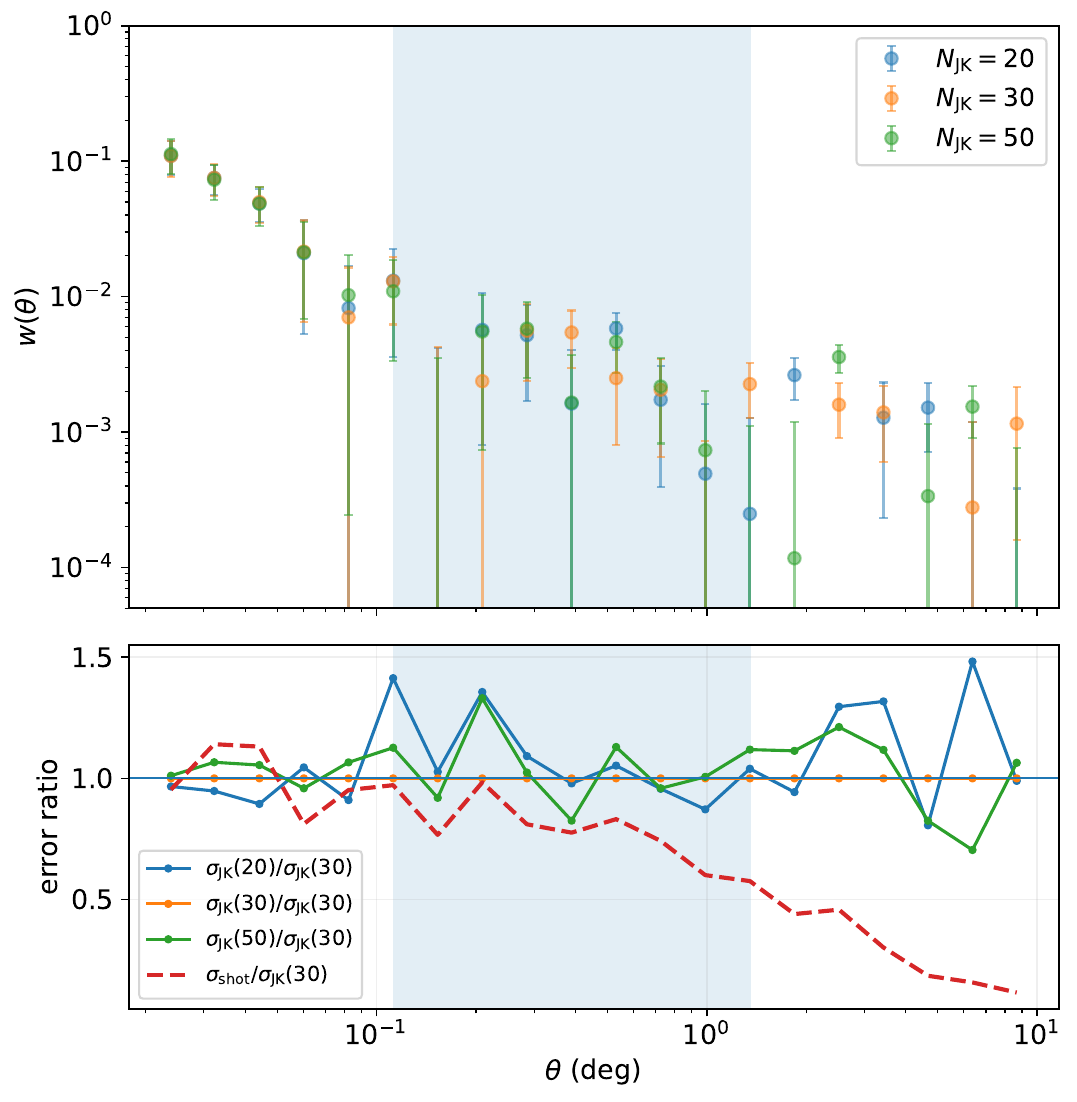}
  \caption{Robustness of \(w(\theta)\) and its uncertainties to the jackknife partitioning. \textbf{Top:} \(w(\theta)\) measured with spatial jackknife covariance estimates using \(N_{\rm JK}=20\), 30 (fiducial), and 50 patches; points show the same binned measurement with the corresponding \(1\sigma\) jackknife errors. \textbf{Bottom:} ratios of the jackknife error amplitudes relative to the fiducial choice, \(\sigma_{\rm JK}(N_{\rm JK})/\sigma_{\rm JK}(30)\), illustrating that the inferred uncertainties are stable to the choice of \(N_{\rm JK}\) over the angular scales used for model fitting (shaded region). The red dashed curve shows the ratio of the purely shot-noise (pair-count) uncertainty estimate to \(\sigma_{\rm JK}(30)\), demonstrating that Poisson errors increasingly underestimate the total uncertainty at large angular separations where sample variance dominates.}
  \label{fig:app_jk_robustness}
\end{figure}

Our fiducial covariance matrix is estimated via spatial jackknife resampling with \(N_{\rm JK}=30\) patches (\autoref{subsec:cov}). Since jackknife uncertainties can depend on how the footprint is partitioned---especially on the largest angular scales where the number of independent modes across the footprint is limited---we test the robustness of both the measured \(w(\theta)\) and the inferred error bars to the choice of \(N_{\rm JK}\).

\autoref{fig:app_jk_robustness} compares the correlation function and jackknife uncertainties obtained using \(N_{\rm JK}=20\), 30 (fiducial), and 50 patches, keeping the sample selection, mask, and binning fixed. The top panel shows that the recovered \(w(\theta)\) measurements are mutually consistent within the uncertainties. The bottom panel quantifies the relative error amplitudes by plotting the ratios \(\sigma_{\rm JK}(20)/\sigma_{\rm JK}(30)\) and \(\sigma_{\rm JK}(50)/\sigma_{\rm JK}(30)\) as a function of \(\theta\). Over the conservative fitting range used for parameter inference (shaded region), the ratios remain close to unity with no systematic trend, indicating that our covariance estimates are not sensitive to moderate changes in the jackknife partitioning scheme. We therefore adopt \(N_{\rm JK}=30\) as a practical compromise: the patches remain large enough to probe variance on degree scales while still providing a sufficient number of resamplings for a stable covariance estimate.

For additional context, we also compare the jackknife uncertainties to a purely pair-count (shot-noise) contribution to the variance (red dashed curve in the bottom panel). As expected for wide-area clustering, this Poisson/shot-noise term becomes increasingly sub-dominant at large separations where sample variance dominates, and therefore underestimates the total uncertainty on degree scales. This motivates our use of the spatial jackknife covariance throughout the main analysis.

\end{document}